\documentclass{article}

\usepackage{syntonly}
\usepackage{arxiv}
\usepackage{comment}
\usepackage[utf8]{inputenc} 
\usepackage[left]{lineno}
\usepackage[T1]{fontenc}    
\usepackage{hyperref}       
\usepackage{url}            
\usepackage{booktabs}       
\usepackage{amsfonts}       
\usepackage{nicefrac}       
\usepackage{microtype}      
\usepackage{amsmath}
\usepackage{lipsum}		
\usepackage{graphicx}
\usepackage{caption}
\usepackage{xcolor}
\usepackage[numbers]{natbib}
\usepackage{doi}
\usepackage{authblk}
\usepackage{amsmath}
\usepackage{bm}
\usepackage{tcolorbox}
\usepackage{soul}  
\usepackage{xcolor} 
\soulregister\cite7  
\soulregister\ref7   

\title{JAX-BTE: A GPU-Accelerated Differentiable Solver for Phonon Boltzmann Transport Equations}

\author[1]{Wenjie Shang}
\author[1]{Jiahang Zhou}
\author[1]{J.P. Panda}
\author[1]{Zhihao Xu}
\author[1]{Yi Liu}
\author[1]{Pan Du}
\author[1,2,*]{Jian-Xun Wang}
\author[1,*]{Tengfei Luo}

\affil[1]{Aerospace and Mechanical Engineering, University of Notre Dame, Notre Dame, IN 46556, USA}
\affil[2]{Mechanical and Aerospace Engineering, Cornell University, Ithaca, NY 14850, USA}
\affil[*]{Corresponding author: \href{mailto:jw2837@cornell.edu}{jw2837@cornell.edu} (Jian-Xun Wang), \href{mailto:tluo@nd.edu}{tluo@nd.edu} (Tengfei Luo)}

\hypersetup{
pdftitle={A template for the arxiv style},
pdfsubject={q-bio.NC, q-bio.QM},
pdfauthor={David S.~Hippocampus, Elias D.~Striatum},
pdfkeywords={First keyword, Second keyword, More},
}

\begin{document}

\maketitle

\begin{abstract}
This paper introduces JAX-BTE, a GPU-accelerated, differentiable solver for the phonon Boltzmann Transport Equation (BTE) based on differentiable programming. JAX-BTE enables accurate, efficient and differentiable multiscale thermal modeling by leveraging high-performance GPU computing and automatic differentiation. The solver efficiently addresses the high-dimensional and complex integro-differential nature of the phonon BTE, facilitating both forward simulations and data-augmented inverse simulations through end-to-end optimization. Validation is performed across a range of 1D to 3D simulations, including complex FinFET structures, in both forward and inverse settings, demonstrating excellent performance and reliability. JAX-BTE significantly outperforms state-of-the-art BTE solvers in forward simulations and uniquely enables inverse simulations, making it a powerful tool for multiscale thermal analysis and design for semiconductor devices.
\end{abstract}


\section{Introduction}
With the rapid advancement of nanotechnology, modern integrated circuits (ICs) can now contain billions of transistors with complex micro- and nanostructures. These densely packed components can generate significant heat in confined spaces, resulting in a large rise in the temperature inside the device, which poses challenges for maintaining reliable and efficient device operation~\cite{garimella2008thermal, prasher2008cooling}. Effective thermal analysis is therefore essential for the design and optimization of ICs to ensure that heat dissipation does not compromise their performance and reliability. 

In scales where diffusive thermal transport described by Fourier's law breaks down, the phonon Boltzmann Transport Equation (BTE) governs heat transfer. Phonons are the quanta of lattice vibration. The importance of BTE lies in its ability to accurately capture thermal transport processes across multiple scales, from the nanoscale, where the ballistic heat transfer dominates, to the macroscale, where diffusive transport holds. This multiscale nature of the phonon BTE is particularly suitable for thermal transport simulations with the dimensions of micro/nano-structures, such as transistors, where the characteristic lengths of these devices approach or even fall below the mean free path of phonons \cite{luo2013nanoscale, minnich2011quasiballistic, cahill2003nanoscale}. Under such conditions, conventional models like Fourier’s law, which assumes a purely diffusive thermal transport mechanism, fail to capture the ballistic nature of thermal phonon transport, leading to inaccuracies in predictions of thermal behavior~\cite{cahill2014nanoscale, chen2001phonon,chen2021non, toberer2012advances, guo2015phonon, minnich2015advances, li2005thermal, majumdar2004thermoelectricity,mazumder2021boltzmann}.

However, solving the phonon BTE poses significant challenges due to its inherently high dimensionality, which involves phonon frequency, polarization, time, spatial coordinates, and directional angles. The nonlinear integro-differential nature of the equation further complicates the solution process, especially when dealing with complex geometries and diverse boundary conditions~\cite{murthy2005review}. Over the years, various numerical phonon BTE solvers have been developed, primarily categorized into stochastic approaches, such as Monte-Carlo methods~\cite{landon2014deviational,mazumder2001monte, lacroix2005monte, mittal2010monte, peraud2011efficient, peraud2015adjoint}, and deterministic discretization-based methods~\cite{escobar2007influence, escobar2008thin, murthy2002computation, narumanchi2004submicron, ali2014large, sheng2024integrating,saurav2023extraction, saurav2024anisotropic, zhang2021fast}. These solvers are predominantly implemented in traditional low-level programming languages such as Fortran or C++, and executed on CPU-based systems to maintain high numerical precision and stability~\cite{bao2018review}. However, despite their accuracy, these traditional solvers suffer from two critical limitations. First, they are often computationally expensive and not scalable. For instance, solving the nongray phonon BTE for micrometer-level transistor simulations with 15-band discretization in wave vector, 128 angular directions, and 880,000 mesh elements using the implicit discrete ordinates method can take up to approximately 1,300 hours on 128 CPU cores in the worst case. Even with algorithmic optimizations and efficient parallel implementations on GPUs, computational costs remain significant~\cite{hu2022ultra}. Techniques such as optimized band discretization~\cite{hu2022optimized} have been proposed to mitigate these costs by reducing the number of discrete bands without sacrificing accuracy, but these methods still face scalability issues for large-scale systems. Second, traditional solvers are typically designed only for forward simulations, lacking support for gradient back-propagation. This severely limits their applicability for inverse problems, which are critical for optimization, parameter estimation, and hybrid neural modeling tasks. Without gradient information, integrating these solvers into gradient-based optimization workflows or neural network models is highly challenging, making it difficult to leverage modern machine learning techniques for end-to-end optimization.


In recent years, physics-informed machine learning (PIML)~\cite{karniadakis2021physics}, particularly physics-informed neural networks (PINNs), have shown promise in solving high-dimensional PDEs such as the phonon BTEs~ \cite{li2021physics,li2022physics,li2023physics,zhou2023physics,zhou2023physics_npj}. By incorporating PDE residuals directly into the loss function of neural networks, PINNs aims to approximate PDE solutions without the need for traditional grid-based discretization methods~\cite{raissi2019physics}. In theory, this approach allows for more efficient handling of complex, high-dimensional problems than conventional numerical solvers. However, despite their success in many applications~\cite{li2021physics,sun2020surrogate,cai2021physics}, PINNs can often face significant challenges in practice, particularly with training stability and convergence.  These issues become especially pronounced in problems involving complex geometries or sharp solution gradients~\cite{wang2022and,gao2021phygeonet}. An alternative PIML approach is to seamlessly integrate ML models with physics-based numerical solvers through differentiable programming, creating neural differentiable models~\cite{liu2024multi,fan2024differentiable,akhare2023diffhybrid}. Unlike PINNs, which rely exclusively on neural networks to approximate the solution, neural differentiable model combines numerical PDE operators with trainable neural networks \cite{liu2024multi}, providing more accurate and physically consistent predictions. This hybrid approach leverages the strengths of both traditional numerical solvers and modern machine learning, allowing for joint optimization of both components, which has shown great promise in a variety of applications~\cite{akhare2023physics,fan2024neural,akhare2024probabilistic}. However, enabling such a unified learning framework that effectively combines these components faces several challenges, especially for phonon BTEs. First, each module (numercial vs. machine learning) is typically developed in different programming languages and relies on distinct computational backends, complicating integration. Second, and crucially, the numerical solver component must be differentiable and highly efficient to enable seamless and scalable optimization, which requires careful design and implementation. These limitations highlight the need for a robust, efficient, and differentiable solver for phonon BTEs. 


In response to the opportunities and challenges discussed above, we developed JAX-BTE, a GPU-accelerated, fully differentiable solver for phonon BTE simulations that seamlessly integrates differentiable programming with high-performance computing. JAX-BTE addresses the need for an efficient, scalable, and differentiable solver by leveraging the capabilities of the JAX framework~\cite{jax2018github, frostig2018compiling}, a high-performance differentiable programming platform for scientific computing and machine learning. Recently, there has been growing interest in differentiable physics, leading to the development of several JAX-based differentiable solvers for applications such as fluid dynamics~\cite{fan2024differentiable, Kochkov2021-ML-CFD,bezgin2024jax}, molecular dynamics~\cite{schoenholz2020jax}, and finite element simulations~\cite{xue2023jax}. However, no existing work has successfully applied differentiable programming to phonon BTE solvers, making JAX-BTE the first of its kind. JAX offers several powerful features, including automatic differentiation (AD), support for diverse backends (CPU, GPU, and TPU), and efficient vectorized operations through Just-in-Time (JIT) compilation. By utilizing these features, JAX-BTE follows this trend by enabling efficient, accurate, and differentiable phonon BTE simulations, paving the way for new capabilities in inverse modeling, sensitivity analysis, and data-driven optimization, a gap that traditional solvers cannot address.
The main advantages of our JAX-BTE over other traditional BTE solvers include:
\begin{enumerate}
    \item Ease of use: JAX-BTE is implemented in Python style, requiring only a few lines of code to set up and run simulations.
    \item High computational efficiency: JAX-BTE is fully vectorized and capable of performing parallel forward simulations on both structured and unstructured meshes, allowing for massive scalability.
    \item Multi-backend support: JAX-BTE supports various computational backends, including CPU and GPU, ensuring flexibility and optimized performance across platforms. \vspace{-0.3em}
    \item Differentiability: Built-in AD enables gradient back-propagation within the solver, allowing to efficiently solve inverse problems and optimization tasks using gradient-based training techniques.\vspace{-0.3em}
\end{enumerate}

The paper is structured as follows. In section "Results", we presents the numerical validation of the JAX-BTE solver, highlighting its AD features and demonstrating its efficiency and reliability through various test cases. Section "Discussion" provides a summary of the key findings and outlines potential directions for future research. Finally, section "Methods" introduces the phonon BTE and provide a detailed discussion of the numerical algorithms implemented in JAX-BTE.

\section{Results}

In this section, we present a comprehensive set of numerical results obtained from our JAX-BTE solver to showcase its forward and inverse simulation capabilities. For validation, we compare our JAX-BTE results with analytical solutions and/or numerical results from existing established solvers (e.g., GiftBTE) for forward simulations in both 1D and 3D geometries. These comparisons demonstrate the accuracy and efficiency of our solver. Additionally, we demonstrate inverse problem-solving capabilities through two example cases, including 1D inverse learning to determine the system dimensions and 2D inverse learning to estimate the heat source intensity. These comprehensive numerical studies highlight the efficiency, accuracy, robustness, and versatility of JAX-BTE in addressing both forward and inverse phonon transport problems.

\subsection{1D cross-plane phonon transport}
\label{sec:1D-forward}
\begin{figure}
	\centering
    \includegraphics[width=0.9\textwidth]{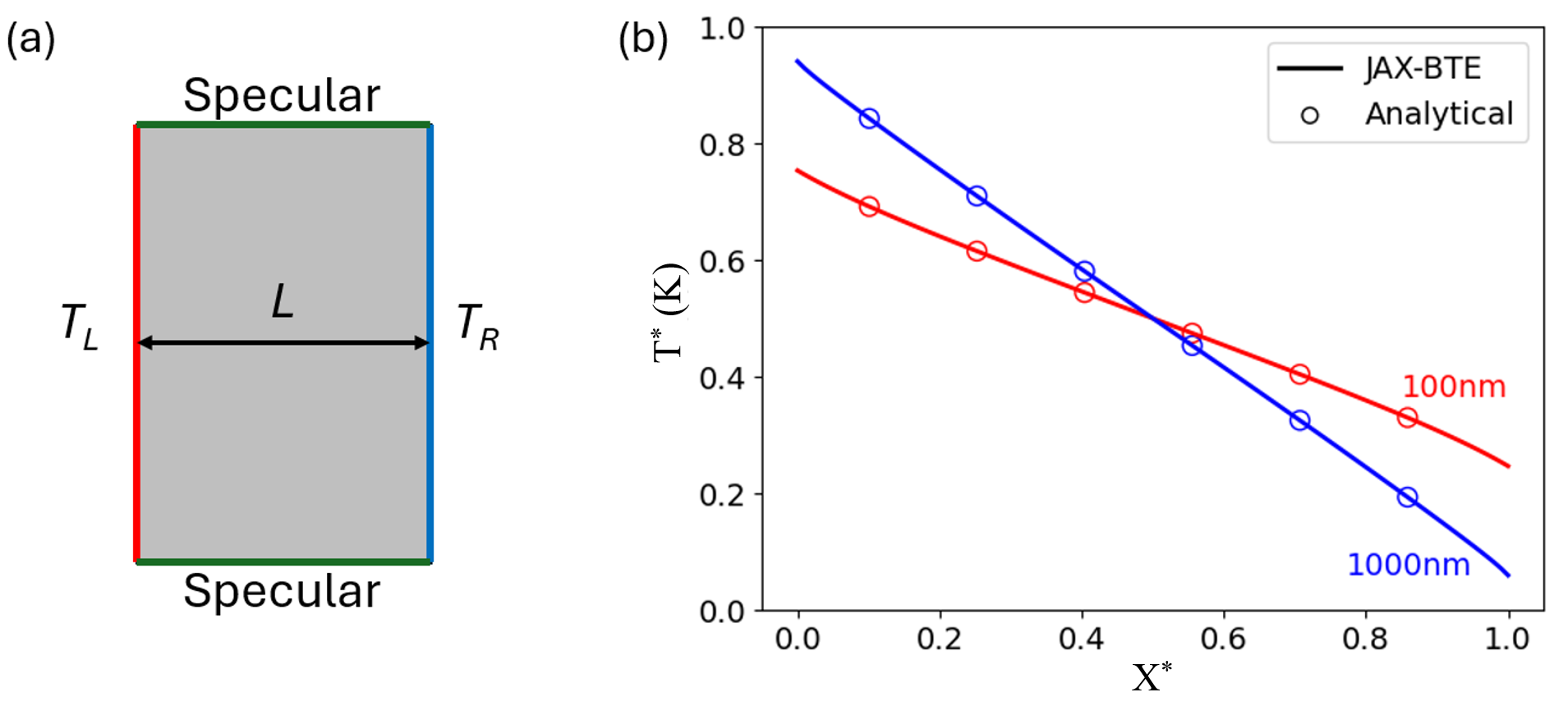}
	\caption{Results of 1D cross-plane phonon transport. (a) The geometry and boundary conditions of the thin films. (b) Dimensionless temperature profiles of the silicon thin films with different thicknesses (L = 100nm, 1000nm), where $T^\ast=(T-T_R)/(T_L-T_R)$ and $X^\ast=x/L$. The solid lines represent the JAX-BTE results and the circles are analytical solutions to the 1D phonon BTE.}
	\label{fig:fig1}
\end{figure}

We first consider heat conduction across a silicon thin film (see Figure~\ref{fig:fig1}a). Given that the lateral dimensions of the film are much larger than the thickness, heat conduction reduces to a 1D problem with two isothermal boundary conditions. The thickness of the film is $L$, and the left and right boundaries are fixed at $T_L=400 $ K and $T_R=300 $ K, respectively.

The spatial domain is discretized into 100 interior cells, and a [4, 4] Gauss-Legendre quadrature for the polar angle $\theta$ and the azimuthal angle $\varphi$ is used for direction discretization in each of the 8 octants in the 3D space. To demonstrate the nonlocal effects on the temperature distributions at varying characteristic dimensions, two plane thicknesses (100 nm, 1000 nm) are studied. Figure~\ref{fig:fig1}b shows the dimensionless temperature profiles $T^\ast=(T-T_R)/(T_L-T_R)$ for the two different thicknesses $L$. The analytical solutions obtained via the method of degenerate kernels are included for comparison~\cite{guo2016discrete,luo2017discrete, sellan2010cross}. The results show that the temperatures predicted by the JAX-BTE solver match with the analytical solutions almost perfectly. For the case with the smaller thickness of 100 nm, the BTE solution captures the temperature ``slip'' near the boundary, a result of non-equilibrium effect.  In contrast, for the larger thickness of 1000 nn, the BTE solution converges to the classical Fourier's heat diffusion behavior.

\begin{figure}
	\centering
    \includegraphics[width=0.55\textwidth]{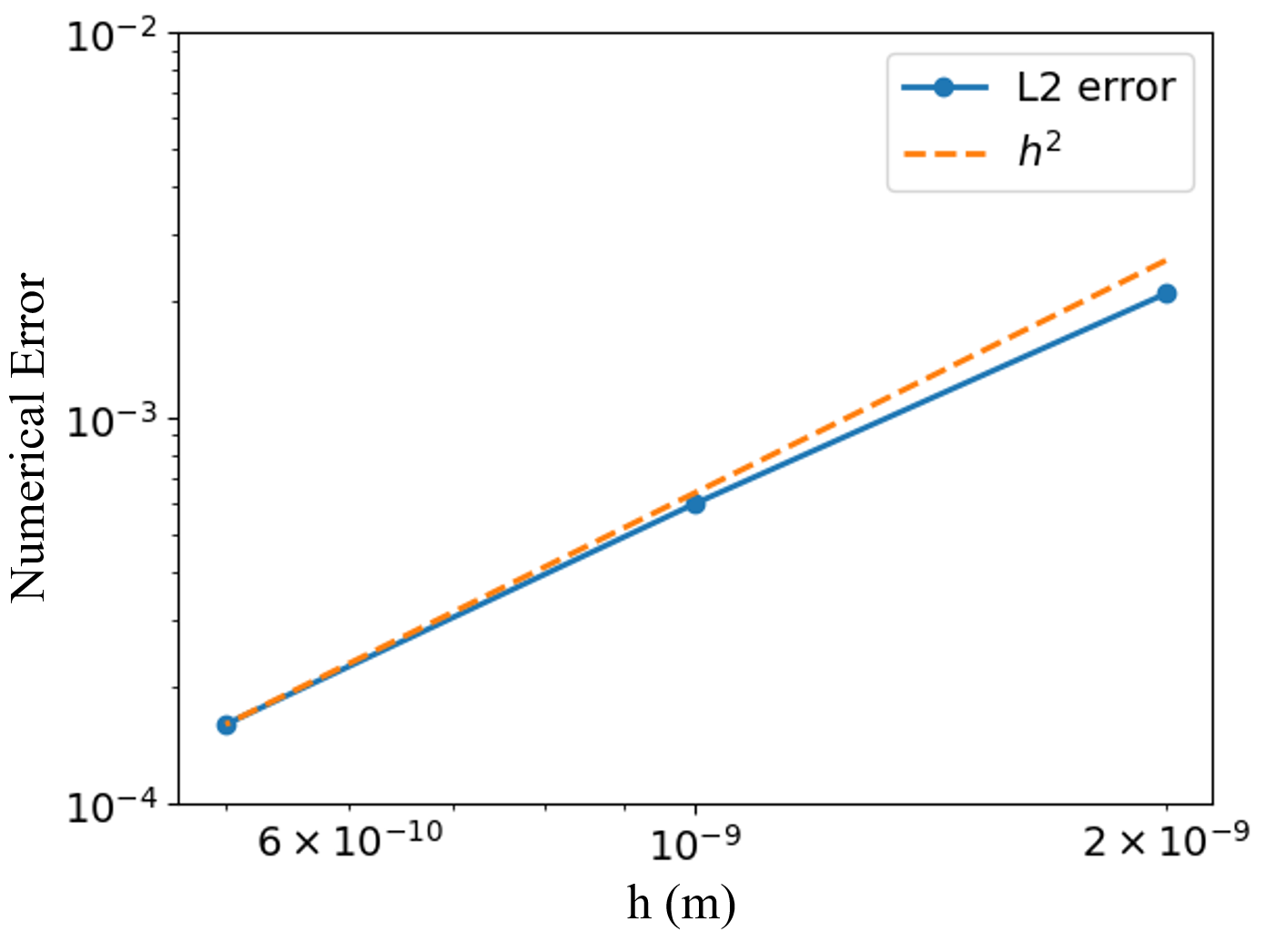}
	\caption{Convergence study for the JAX-BTE solver. The plot shows the numerical error (L2 error) as a function of grid size on a log-log scale for a 1D case. A reference line (orange) representing $h^2$ vs. $h$ is included to indicate the expected second-order accuracy of the numerical method. The observed error line (blue) calculated by comparing the JAX-BTE solver results with the semi-analytical solution using three different meshes (50, 100, and 200 cells).}
	\label{convergence}
\end{figure}

To verify the numerical accuracy of JAX-BTE, we conducted a grid convergence study, comparing the solver's numerical error against a semi-analytical solution for the 1D cross-plane case [66]. Figure~\ref{convergence} presents the discretization error (L2 error) as a function of grid size on a log-log scale. A reference line representing the formal second-order accuracy ($h^2$ vs. $h$) is included for comparison. The observed order of accuracy closely matches the formal order of accuracy, as the error curve aligns well with the reference slope, confirming that JAX-BTE achieves near-second-order convergence. 

Specifically, for grid sizes of $2\times10^{-9}$ m, $1\times10^{-9}$ m, and $5\times10^{-10}$ m, the corresponding L2 errors are $2.1\times10^{-3}$, $6\times10^{-4}$, and $1.6\times10^{-4}$, respectively,  yielding observed orders of accuracy of 1.8 and 1.9, which are close to the theoretical second-order accuracy. The slight deviation from the expected value of 2.0 is attributed to numerical approximations arising from iterative solvers, angular and band discretization, and the semi-analytical reference solution. Despite these small discrepancies, the observed convergence rate demonstrates that JAX-BTE provides numerically consistent and accurate solutions, further validating its reliability for phonon transport simulations.


\subsection{3D FinFET transistor with long base}

\begin{figure}
	\centering
    \includegraphics[width=0.85\textwidth]{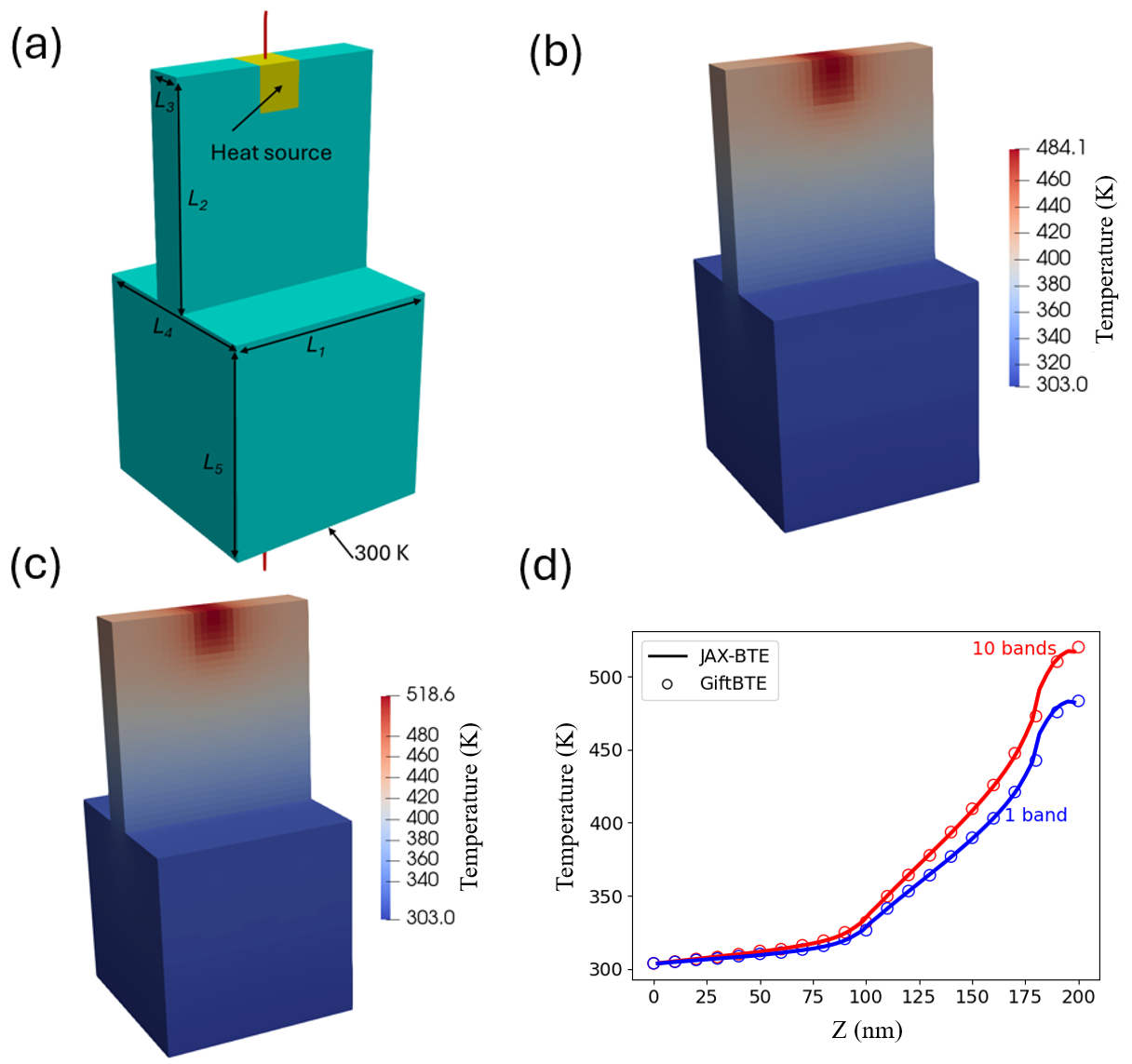}
	\caption{Results of 3D FinFET simulations with diffusive reflecting boundary conditions of all faces except the bottom surface, which is set at 300 K as an isothermal boundary condition. (a) The structure and the computational domain of the FinFET. The yellow part is the heat source region. The red axis in the plot shows the path, along which the temperature profile was extracted (b) Temperature field of FinFET predicted by the JAX-BTE using 1 phonon band (i.e., gray model). (c) Temperature field of FinFET predicted by the JAX-BTE using 10 discreteized bands in the wave vector domain (i.e., non-gray model). (d) Temperature profiles extracted from the center line of the structure along the z-axis. The solid lines represent the JAX-BTE results and circles are the GiftBTE solutions.}
	\label{fig:fig2}
\end{figure}

We select the fin field-effect transistor (FinFET) as a representative example of transistors and used JAX-BTE to predict its temperature distribution. The structure and boundary conditions of the FinFET are illustrated in Figure~\ref{fig:fig2}a, with dimensional details provided in Table~\ref{dimension3.2}. The bottom thermalizing boundary is maintained at 300 K to simulate a room temperature environment. A diffusely reflecting boundary (i.e., adiabatic boundary) is applied to all boundaries except the bottom wall. To evaluate the performance of the JAX-BTE solver, we selected band numbers 1 and 10 to assess both the gray and non-gray scenarios. The hot spot is located in a cubic region near the top surface with a uniformly distributed heat source. The volumetric heat source intensity is set at 10\textsuperscript{19} W/m\textsuperscript{3}, which mimics a hot spot in a transistor due to Joule heating~\cite{maiti2017introducing, hao2018hybrid}.

After conducting a convergence test, we utilized 128 directions ($4\times4 \times 8$) in the solid angle space and a structured mesh consisting of 36,085 hexahedral elements. For validation, the same settings were used for both our JAX-BTE solver and the widely recognized high-preformance C++ solver, GiftBTE~\cite{hu2023giftbte}, to obtain the temperature distributions. As shown in Figure~\ref{fig:fig2}d, the temperature profiles extracted along the centerline of the structure in the $z$ direction from both JAX-BTE and GiftBTE exhibit a perfect match. This agreement between the two solvers demonstrates the reliability and accuracy of our JAX-BTE implementation in capturing the thermal transport in complex geometries.

\begin{table}
	\caption{Dimension of long-base FinFet's edges}
	\centering
	\begin{tabular}{c c c c c c}
		\toprule
		Location     & $L_1$     & $L_2$ & $L_3$  & $L_4$  & $L_5$ \\
  	\midrule
		Length (nm) & 100  & 100 & 30 & 100 & 100     \\
		\bottomrule
	\end{tabular}
	\label{dimension3.2}
\end{table}

\subsection{3D FinFET transistor with short base}

\begin{figure}[htp!]
	\centering
    \includegraphics[width=0.85\textwidth]{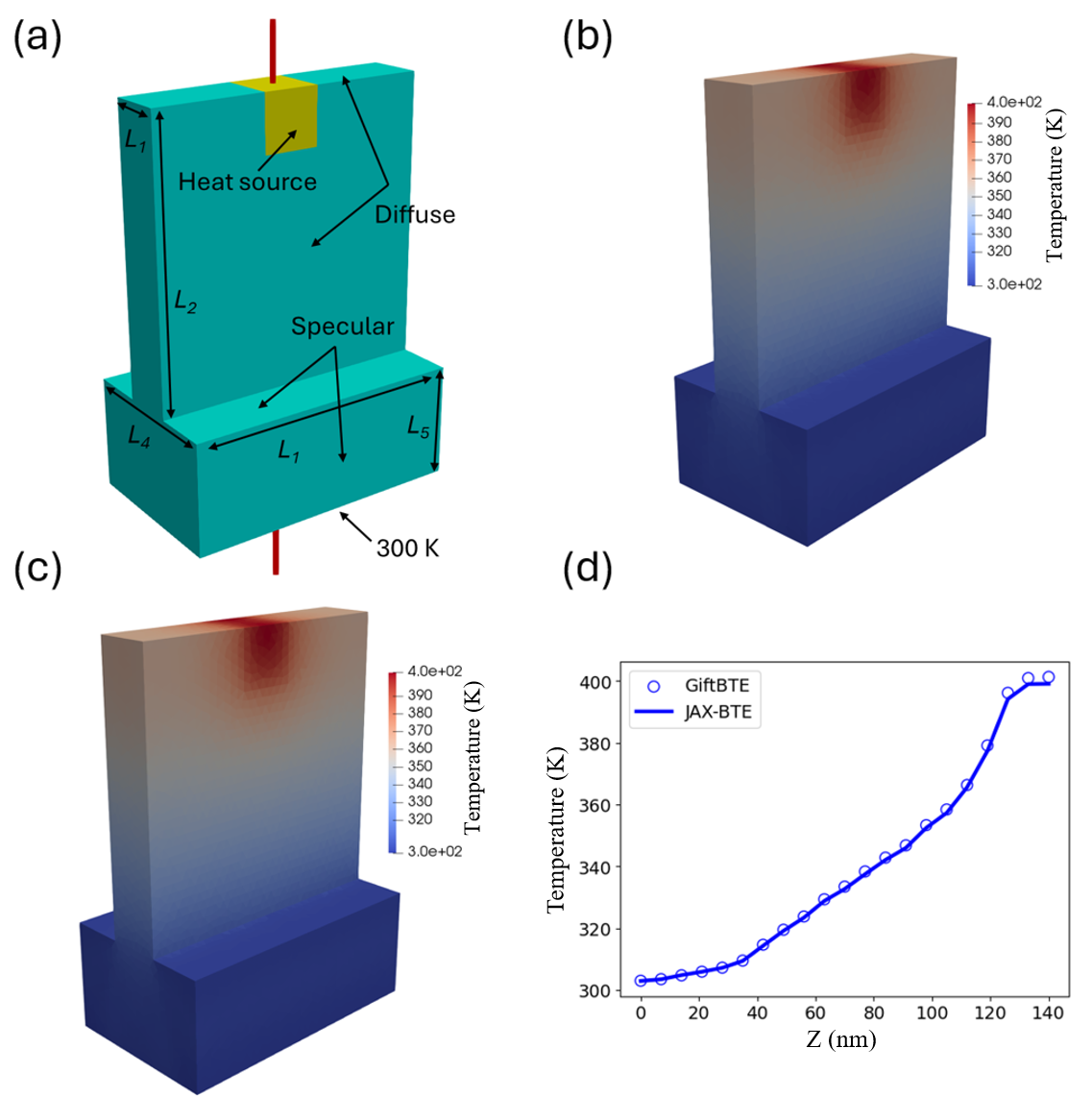}
	\caption{Results of 3D FinFET simulations with specularly and diffusely reflecting boundary conditions. (a) The structure and the computational domain of the FinFET. The red axis in the plot represents the path, along which the temperature profile was extracted. (b) Temperature field of the FinFET predicted by the JAX-BTE. (c) Temperature field of the FinFET predicted by GiftBTE. (d) Temperature profiles extracted from the center line along the vertical axis. The solid lines represent the JAX-BTE result and circles are the GiftBTE result.}
	\label{fig:fig3}
\end{figure}

To further validate the JAX-BTE solver, we simulate another FinFET structure with a combination of specularly and diffusely reflective boundary conditions. The structure and boundary conditions of the FinFET are illustrated in Figure~\ref{fig:fig3}a, with dimension details provided in Table \ref{dimension3.3}. Key boundary conditions include a thermalizing boundary set at 300 K to simulate contact with metal electrodes on top of the fin and the substrate at the bottom, diffusely reflecting boundaries for fin surfaces and specularly reflecting boundary condition for all base surfaces, as shown in Figure~\ref{fig:fig3}a. The heat source is located near the top surface as a square box, with a fixed heat source intensity to mimic heat generation during transistor operations.

\begin{table}[htp!]
	\caption{Dimension of short-base transistor's edges}
	\centering
	\begin{tabular}{c c c c c c }
		\toprule
		Location     & $L_1$  & $L_2$ & $L_3$  & $L_4$  & $L_5$  \\
  	\midrule
		Length (nm) & 100  & 100 & 20 & 60 & 40    \\
		\bottomrule
	\end{tabular}
	\label{dimension3.3}
\end{table}

The FinFET structure is discretized using an unstructured mesh with 46,075 cells, based on the mesh convergence test. A $2\times4\times8$ angle discretization strategy is employed for direction discretization, and four phonon bands are used to approximate different phonon states. A uniform heat source with intensity of 5$\times$10\textsuperscript{18} W/m\textsuperscript{3} is applied as the heat generation term. Figure~\ref{fig:fig3}d shows the temperature profile along the vertical axis extracted from the JAX-BTE simulation results, along with the GiftBTE-simulated temperature profile. It is observed that the JAX-BTE result agrees well with the GiftBTE-simulated temperature profile. This excellent agreement demonsrates the reliability and accuracy of the JAX-BTE solver in capturing the thermal distributions under complex boundary conditions.

\subsection{3D FinFET transistor array with long base}
    
\begin{figure}[htp!]
	\centering
    \includegraphics[width=0.85\textwidth]{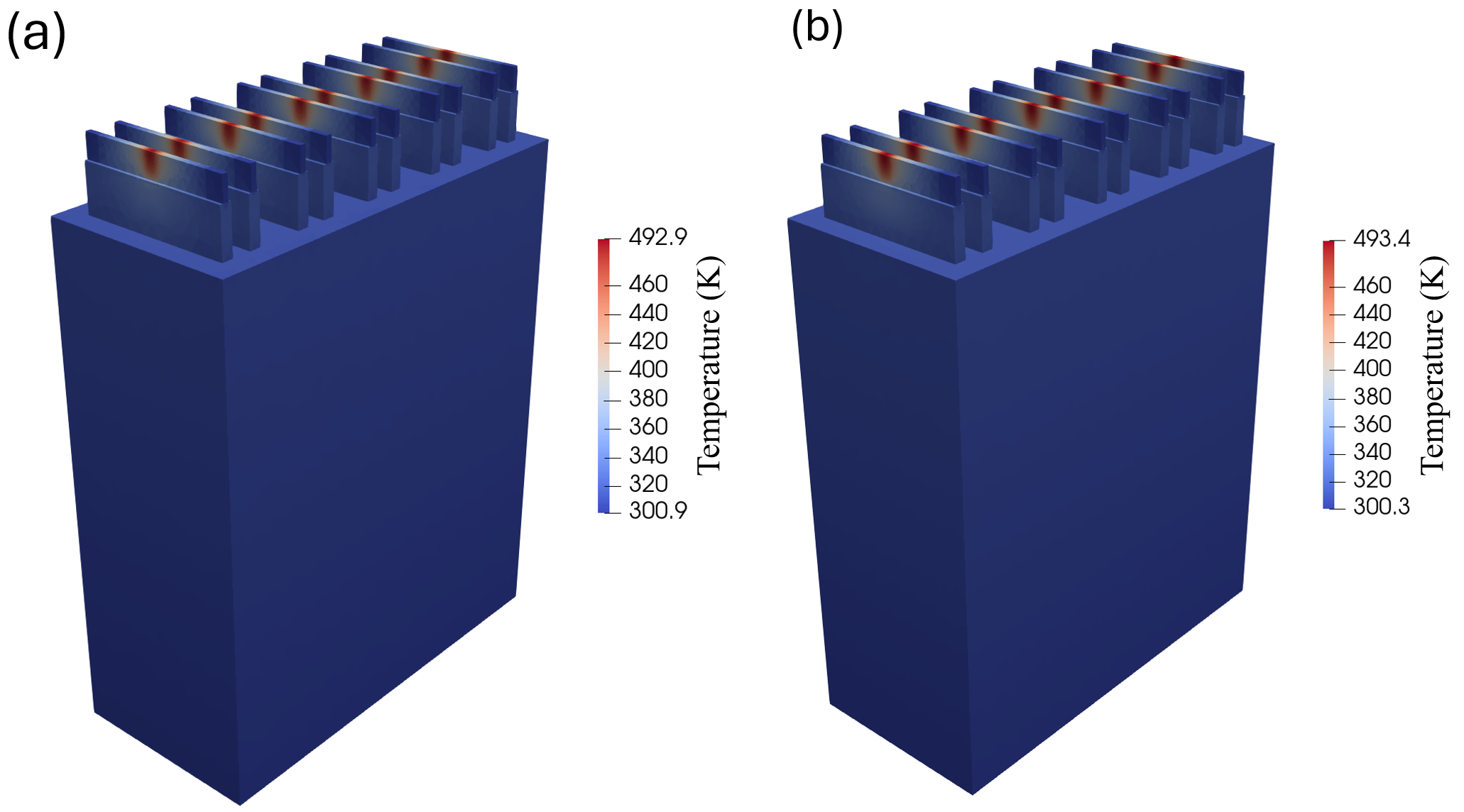}
	\caption{Temperature distribution for the linear transistor array simulation. (a) JAX-BTE results. (b) GiftBTE results. }
	\label{test4}
\end{figure}

To further assess the robustness of our JAX-BTE for real-world applications, we simulated a linear array of five transistors, each containing two fins, placed side by side on a 700 nm silicon base. The computational domain was discretized into 197,500 cells with a single phonon band and $2 \times 2$ angular discretization. The bottom surface of the domain was set to 300 K, while the top surfaces of the transistor fins were assigned a combination of isothermal and adiabatic boundary conditions to closely mimic the real operating applications. All other boundaries were modeled to be specularly reflective.

A uniform heat source with an intensity of 10\textsuperscript{19} W/m\textsuperscript{3} was applied to the transistor fins, simulating an ``on'' state for all transistors. The temperature distributions obtained using JAX-BTE (Figure~\ref{test4}a) and the benchmark solver, GiftBTE (Figure~\ref{test4}b), exhibited excellent agreement. The maximum temperatures predicted by JAX-BTE and GiftBTE were 492.9 K and 493.4 K, respectively, with a difference of only $1\%$. This close match validates the accuracy and reliability of JAX-BTE in accurately capturing multiscale thermal transport within realistic microelectronic structures.

This test case further highlights the capability of JAX-BTE to handle large-scale, realistic geometries with complex boundary conditions, demonstrating its robustness and applicability for practical thermal transport simulations.

\subsection{1D inverse problem for system size}

Inverse problems involve determining unknown parameters of a system from observable data. Our JAX-BTE solver, leveraging the differentiable programming capabilities of JAX, enables end-to-end optimization for such tasks.

In this case study, we focus on heat conduction in a 1D silicon thin film, as described in 1D cross-plane phonon transport, with the only difference being that the film thickness $L$ is unknown. The goal of the inverse problem is to determine the unknown film thickness $L$ that results in a given temperature profile. Temperature observations are taken at 20 uniformly spaced points along the domain, and the L2 norm is used to measure the discrepancy between the simulated and observed temperatures.

\begin{figure}[htp!]
	\centering
    \includegraphics[width=0.9\textwidth]{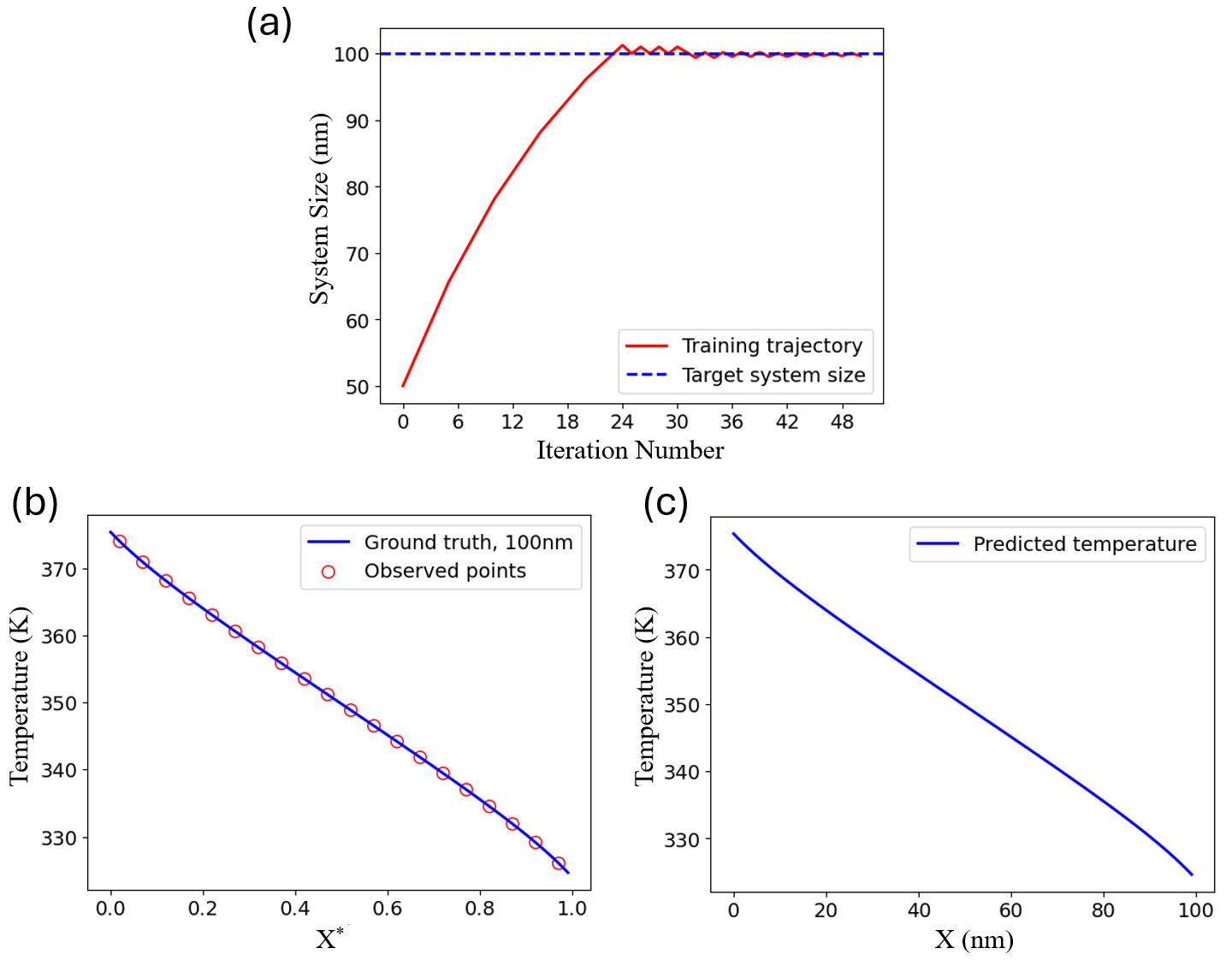}
	\caption{Results of inverse learning for the system size of the 1D thin film. (a) The predicted system size trajectory during the inverse learning iterations. The dashed blue line represents the 100 nm ground truth system size. (b) The temperature profile with ground truth system size of 100 nm. Red circles are temperatures at the observed points used for the JAX-BTE inverse learner. (c) The predicted temperature profile using the observed data.}
	\label{fig:fig4}
\end{figure}

The film thickness is parameterized and incorporated into the mesh initialization process for inverse learning. Starting with an initial guess of 50 nm, we simulate the temperature profile using JAX-BTE. The discrepancy between this simulated temperature and the target temperature distribution from the ground truth setup is quantified using the L2 norm between the predicted and observed temperatures at the 20 selected points. The differentiability of JAX-BTE enables the use of gradient descent to iteratively adjust the system size $L$ by minimizing the L2 loss. An exponential decay scheduler is applied to improve the convergence rate. Within 25 iterations, the process converges to the correct film thickness of 100 nm. Figure~\ref{fig:fig4}a shows the converging of the system size over iterations.

These results demonstrate that JAX-BTE, through the differentiable programming, facilitates efficient and accurate inverse design. The solver's ability to accurately converge to the true system size validates its effectiveness and robustness in practical applications of inverse design in nanoscale heat transfer. This approach highlights the potential of JAX-BTE in broadening the scope of computational experiments for device design and material selection.

\subsection{2D inverse problem for heat source intensity}

\begin{figure}[htp!]
	\centering
    \includegraphics[width=0.9\textwidth]{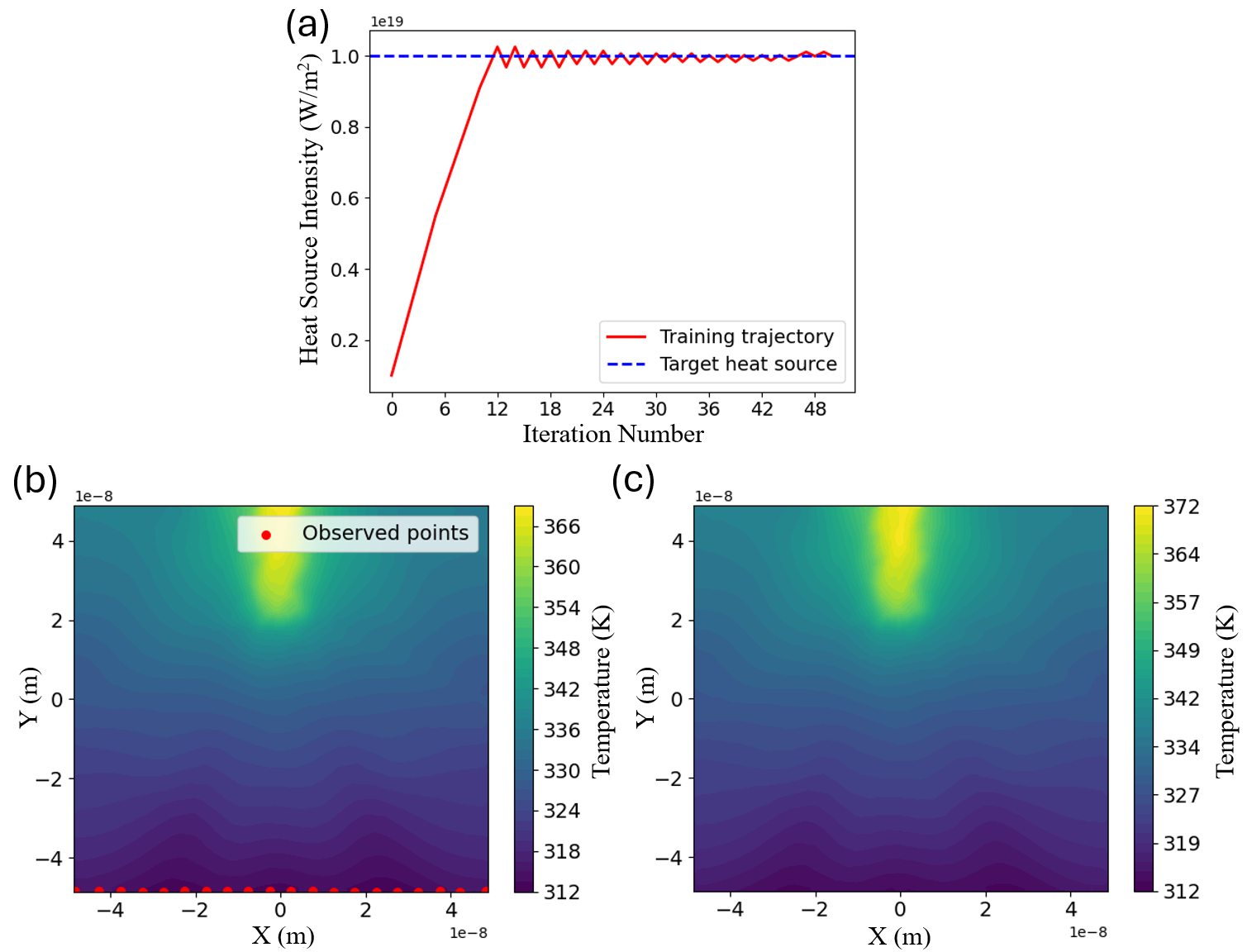}
	\caption{Results of inverse learning for the heat source intensity of the 2D square domain. (a) The predicted heat source intensity trajectory during the inverse learning iterations. The dashed blue line represents the ground truth heat source intensity. (b) The temperature distribution with ground truth heat source intensity, 10\textsuperscript{19} W/m\textsuperscript{2}. Red dots are temperatures at the observed points for the JAX-BTE inverse learner. (c) The predicted temperature distribution using the observed data.}
	\label{2dinv}
\end{figure}

In this example, we address the inverse problem of predicting the heat source intensity within a 2D domain using sparse temperature observations. The boundary conditions are set to be isothermal at 300 K on the bottom and specularly reflective on all other sides. The domain is a 100 nm $\times$ 100 nm square with a heat source located in a 10 nm $\times$ 20 nm rectangular box close to the top surface. The domain is discretized using an unstructured mesh with approximately 1,000 cells. The true heat source is uniformly distributed within the box, with an intensity set at 10\textsuperscript{19} W/m\textsuperscript{2}. The ground truth temperature distribution is obtained by solving the phonon BTE using the JAX-BTE solver. Temperature observations are taken at 20 evenly spaced locations along the bottom face (Figure~\ref{2dinv}b). It is important to note that in the JAX-BTE inverse solver, the temperature at locations other than these 20 points remains unknown. 

Following the same procedure as in the 1D case, we begin with an initial guess for the heat source intensity of 10\textsuperscript{18} W/m\textsuperscript{2} and use gradient descent to minimize the discrepancy between the predicted and observed temperatures. After 10 iterations, the predicted heat source intensity converges to the ground truth value. The rapid convergence in this inverse problem demonstrates the effectiveness of our JAX-BTE solver in accurately determining heat source intensities from sparse temperature observations. This case has direct practical implications, as quantifying the heat source in transistors is often challenging, whereas temperature measurements are more readily obtainable. The ability to infer heat source intensity using the JAX-BTE framework could significantly enhance the accuracy of thermal management in nanoelectronics.

\subsection{Performance analysis}

To comprehensively evaluate the performance of JAX-BTE, we conducted a scalability study across varying problems, ranging from approximately 1 million to 110 million degrees of freedom (DoF). Both JAX-BTE and GiftBTE employed the BiCG solver with identical settings, running 300 iterations for each case to ensure a fair comparison. The JAX-BTE simulations were tested on two type of GPUs (NVIDIA RTX 4090 and A100), while GiftBTE was executed on an AMD EPYC 7543 CFD system, utilizing 8, 16, 32, and 64 CPU cores.

\begin{figure}[htp!]
	\centering
    \includegraphics[width=0.65\textwidth]{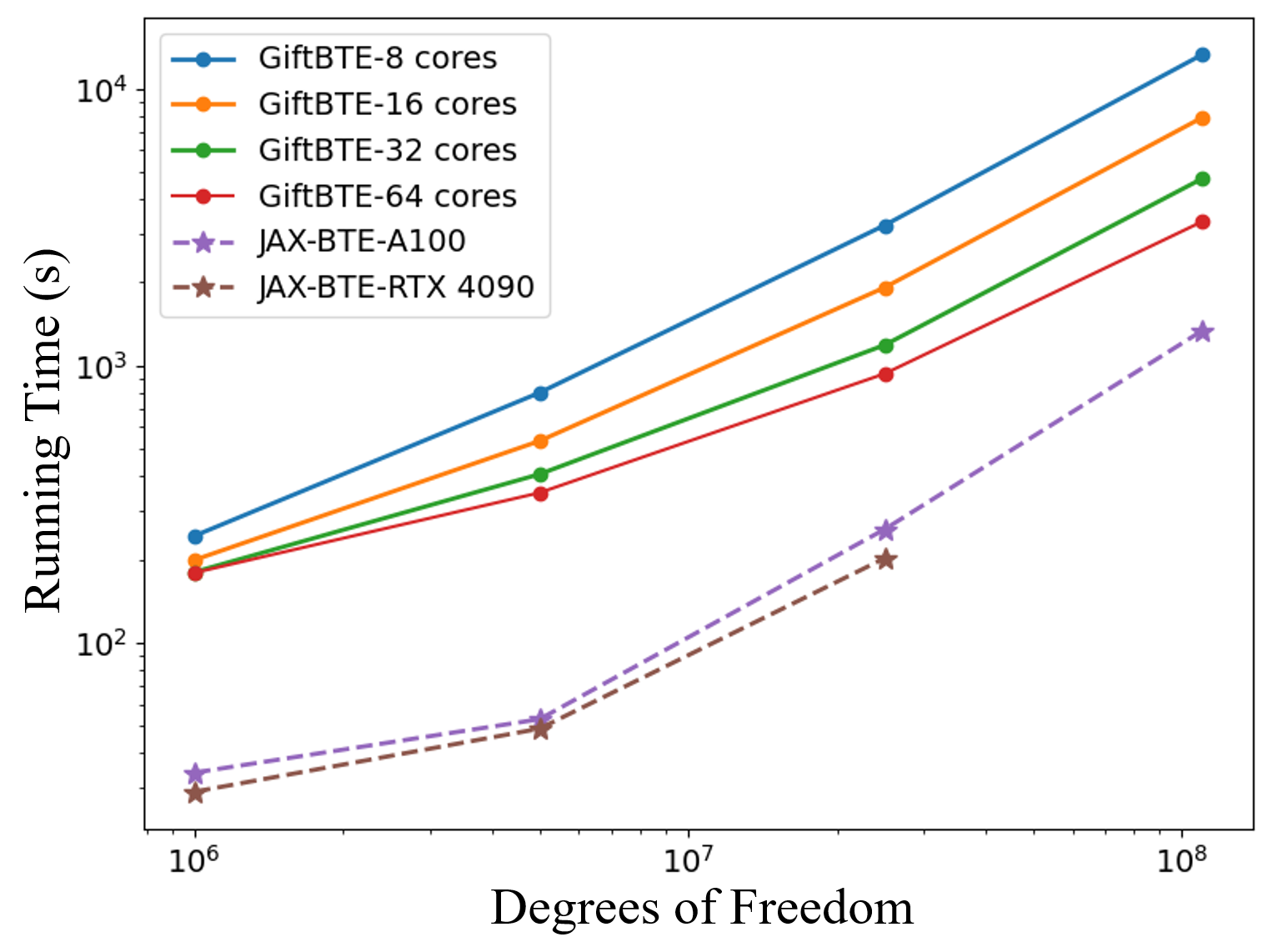}
	\caption{Scalability comparison between JAX-BTE and GiftBTE across different problem sizes, ranging from $10^6$ to $10^8$ DoF.JAX-BTE was executed on single GPUs (RTX 4090 and A100), while GiftBTE was tested on an AMD EPYC 7543 CPU system using 8, 16, 32, and 64 cores.}
	\label{scalability}
\end{figure}

The results, presented in Figure~\ref{scalability}, demonstrate that JAX-BTE consistently outperforms GiftBTE across all problem sizes, exhibiting superior scalability and computational efficiency.For the smallest case (1M DoF),, JAX-BTE completes the computation in just 29 seconds on an RTX 4090 and 34 seconds on an A100 GPU, whereas GiftBTE requires 243 seconds on 8 CPU cores, 199 seconds on 16 cores, and 179 seconds on 64 cores, achieving a 6–8× speedup over GiftBTE.

As the problem size increases, JAX-BTE maintains a significant performance advantage. For the 25M DoF case, JAX-BTE completes the simulation in 202 seconds (RTX 4090) and 257 seconds (A100), compared to 936 seconds for GiftBTE on 64 CPU cores, resulting in a 4x acceleration. For the largest tested case (110M DoF), JAX-BTE on the A100 completes the simulation in 1340 seconds, while GiftBTE requires 3317 seconds even with 64 CPU cores. The RTX 4090, while highly efficient for smaller cases, was unable to run this simulation due to GPU memory limitations.

It is important to emphasize that JAX-BTE achieves this performance using a single GPU, while GiftBTE relies on multi-core CPU parallelization. Scaling JAX-BTE to multiple GPUs would further enhance its efficiency, enabling even greater speedups for large-scale phonon transport simulations.

\section{Discussion}
This study demonstrates the reliability and efficiency of the GPU-accelerated, differentiable solver, JAX-BTE, for nanoscale heat transfer problems. The validation cases have been rigorously designed to assess the accuracy and reliability of JAX-BTE across a variety of scenarios. The accuracy has been validated through comparisons with analytical solutions and other established numerical solvers. Moreover, the differentiability of JAX-BTE has been demonstrated through 1D and 2D inverse problems using sparse temperature measurements. These examples not only affirm the solver’s accuracy in forward settings but also highlight its potential to facilitate material and structural design in inverse settings. Unlike conventional solvers that lack built-in differentiability, JAX-BTE enables automatic gradient computations, making it uniquely suited for sensitivity analysis, inverse learning, and machine learning-assisted optimization. By seamlessly integrating differentiable programming with high-performance computing, JAX-BTE bridges the gap between numerical phonon transport simulations and modern AI-driven engineering workflows.

The promising AD ability of our solver open new avenues for its application in future inverse design tasks. By integrating JAX-BTE with experimental data and optimization techniques, we can enhance the design and development of micro and nanoscale electronic devices. The solver's ability to handle multiple scales and complexities in thermal simulations makes it a versatile tool for researchers and engineers working at the forefront of semiconductor technology and materials science.

However, the current implementation of JAX-BTE is constrained by GPU memory, limiting the maximum DoF the solver can handle to approximately 100 million when using a single NVIDIA A100 GPU card. For more complex BTE simulations and inverse learning tasks, higher spatial and angular resolutions may be necessary, potentially exceeding this limit. To overcome these limitations, future work will focus on developing advanced strategies to expand the solver’s capabilities. Specifically, we aim to implement multi-GPU parallelism and out-of-core computation techniques to distribute data and computation efficiently across multiple GPUs. These enhancements will significantly increase the solver's capacity, enabling it to handle even larger BTE simulations. Additionally, incorporating adaptive mesh refinement and advanced preconditioning techniques could further optimize performance and memory usage.

Furthermore, we plan to extend JAX-BTE by incorporating interface models to support simulations involving multi-material coupling and by incorporating transient analysis capabilities~\cite{dames2004theoretical, song2021evaluation}. These enhancements will enable simulations of complex, multi-material systems and dynamic thermal processes, broadening the applicability of JAX-BTE to a wider range of practical engineering problems.

By addressing these challenges, JAX-BTE will not only enhance its applicability to more complex scenarios but also push the boundaries of what can be achieved in nanoscale heat transfer simulations and inverse design. These advancements will be essential for driving innovation in the next generation of semiconductor devices and advanced materials.

\section{Methods}
\subsection{Phonon BTE}
The phonon BTE describes the evolution of phonon distributions over time in a system. For a system with a heat source at steady state, the energy-based phonon BTE under the single mode relaxation time approximation can be expressed as~\cite{ziman2001electrons, loy2013fast}:
\begin{equation}
    \bm{v} \cdot\nabla e=\frac{e-e^0}{\tau}+Q, 
\end{equation}
where $e(\bm{x}, \bm{s}, \omega, p)$ is the phonon energy distribution function, which depends on spatial position $\bm{x}$, directional unit vector $\bm{s}$, phonon frequency $\omega$, and phonon branch $p$. $\bm{v}(\omega, p)$ is the group velocity, and $\tau(\omega, p)$ is the relaxation time, which can be obtained for all phonons at different $\omega$ and $p$ using first-principles Density Functional Theory (DFT) calculations. $Q$ represents the volumetric heat source intensity, and $e^0(\bm{x})$ is the equilibrium part of the distribution function. According to energy conservation, $e^0$ is related to $e$ through~\cite{bhatnagar1954model}:
\begin{gather}
            e^0 =\frac{1}{4\pi}CT_L,\\
        T_L =\frac{\int_{4\pi}\sum_{p}\int_{\omega}{\frac{e}{\tau}d\omega d\Omega}}{\sum_{p}\int_{\omega}{\frac{C}{\tau}d\omega}},
        \label{update_e0}
\end{gather}
where $C(\omega, p)$ is the volumetric heat capacity for a given phonon mode, $T_L$ is the lattice temperature, and $\Omega$ is the solid angle in spherical coordinates. The heat flux can be calculated by:
\begin{equation}
    \mathbf{q} =\int_{4\pi}\sum_{p}\int_{\omega}\mathbf{v} e d\omega d\Omega.
\end{equation}

\subsection{Boundary conditions}
For the JAX-BTE solver, three built-in boundary conditions commonly used in phonon BTE simulations \cite{guo2016discrete, luo2017discrete} are implemented.

\textit{Isothermal Boundary:} This boundary absorbs all incoming phonons and emits phonons in thermal equilibrium at a specific temperature $T$ back into the simulation domain. This can be described as:
\begin{equation}
    e\left(\bm{x}_b,\bm{s},\omega,p\right)=\frac{C}{4\pi}T,\ \bm{s}\cdot \bm{n}_b<0,
\end{equation}
where $\bm{x}_b$ is the position at the boundary and $\bm{n}_b$ is the outward-pointing normal unit vector at the boundary.

\textit{Diffusely Reflecting Boundary:} This adiabatic boundary reflects phonons leaving the simulation domain with equal probability in all possible directions pointing into the simulation domain. According to energy conservation, it is represented as:
\begin{equation}
    e\left(\bm{x}_b,\bm{s},\omega,p\right)=\frac{1}{\pi}\int_{\bm{s}^\prime\cdot \bm{n}_b>0} e\left(\bm{x}_b,\bm{s}^\prime,\omega,p\right) \left|\bm{s}^\prime\cdot \bm{n}_b\right| d\Omega,\ \bm{s}\cdot \bm{n}_b<0.
\end{equation}
where $\bm{s}^\prime$ are all angles that point out of the domain and the use of this boundary condition implies a rough surface.

\textit{Specularly Reflecting Boundary:} Another adiabatic boundary condition, specular reflecting boundaries assumes phonons reflect as if from a mirror, where the phonon energy along the reflected angle is equal to that of the incident angle:
\begin{equation}
    e\left(\bm{x}_b,\bm{s},\omega,p\right)=e\left(\bm{x}_b,\bm{s}^\prime,\omega,p\right),\ \bm{s}\cdot \bm{n}_b<0,
\end{equation}
where $\bm{s}^\prime=\bm{s}-2\bm{n}_b(\bm{s}\cdot \bm{n}_b)$ is the reflected phonon direction. This boundary condition assumes a perfectly smooth surface.

\subsection{Band discretization}
For discretizing phonon energy in the frequency and polarization domains, the wave vector space is divided into $n$ equidistant intervals $[K_0,\ \ldots,\ K_n]$, referred to as bands, where $K_0$ is the minimum wave vector and $K_n$ is the maximum wave vector. For each wave vector interval, phonon properties such as heat capacity, relaxation time, and group velocity are calculated by weighted averaging these properties within the interval (i.e., within each band). Specifically, for each band, the representative phonon properties are given by~\cite{hu2023giftbte, hu2020unification}:
\begin{gather}
    C_\lambda=\sum_{k>K_{n-1}}^{K_n}C, \\
    \bm{v}_\lambda=\frac{\sum_{k>K_{n-1}}^{K_n}C\bm{v}}{\sum_{k>K_{n-1}}^{K_n}C}, \\
    \tau_\lambda=\frac{\sum_{k>K_{n-1}}^{K_n}{Cv^2}\tau}{\bm{v}_\lambda\sum_{k>K_{n-1}}^{K_n}Cv},\\
    Q_\lambda = \frac{C_\lambda Q}{\sum_\lambda C_\lambda},
\end{gather}
where $\lambda$ is the index of the phonon band and $k$ is the wave vector. The original integral equation of frequency is transformed into a summation form:
\begin{gather}
    \sum_{p}{\int_{\omega}\frac{e}{\tau}d\omega=\sum_{\lambda}\frac{e_\lambda}{\tau_\lambda}\ }, \\
    \sum_{p}{\int_{\omega}\frac{C}{\tau}d\omega=\sum_{\lambda}\frac{C_\lambda}{\tau_\lambda}\ }.
\end{gather}

\subsection{Angle discretization}
The angle discretization transforms the integration over directional space into a weighted summation~\cite{li2021physics, guo2016discrete}:
\begin{equation}
    \int_{4\pi}{e_\lambda d\Omega=\int_{0}^{\pi}\int_{0}^{2\pi}{e_\lambda s i n\theta d\theta d\varphi}}=\sum_{\alpha}{\beta_{\theta_\alpha}\beta_{\varphi_\alpha}sin\theta_\alpha}e_{\alpha,\lambda}=\sum_{\alpha}{\beta_\alpha e_{\alpha,\lambda}}.
\end{equation}
Here $\theta$ is the polar angle and $\varphi$ is the azimuthal angle. $\alpha$ is the index for the sampled direction $\bm{s}_\alpha=(cos\theta,sin\theta cos\varphi,sin\theta sin\varphi)$. The weights $\beta_{\theta_\alpha}$ and $\beta_{\varphi_\alpha}$, as well as their corresponding directions, are calculated using Gauss-Legendre quadrature to achieve high numerical accuracy with reduced computational cost. The final weight of the phonon energy in direction $\alpha$ and band $\lambda$, $e_{\alpha,\lambda}$, is given as $\beta_\alpha=\beta_{\theta_\alpha}\beta_{\varphi_\alpha}sin\theta_\alpha$.

\subsection{Spatial discretization}
After applying band and angular discretization, the phonon BTE in its steady-state form becomes:
\begin{equation}
    \bm{v}_\lambda\cdot\nabla e_{\alpha,\lambda}=-\frac{e_{\alpha,\lambda}-e_\lambda^0}{\tau_\lambda}+Q_\lambda
    \label{eq:bte-2}
\end{equation}
where $\lambda$ is the index of the band and $\alpha$ is the angle index. The equilibrium energy is represented by:
\begin{equation}
\begin{aligned}
    e_\lambda^0 &= \frac{1}{4\pi}C_\lambda T_L, \\
    T_L &= \frac{\sum_{\lambda}\sum_{\alpha}{\omega_\alpha\frac{e_{\alpha,\lambda}}{\tau_\lambda}}}{\sum_{\lambda}\frac{C_\lambda}{\tau_\lambda}}.
    \label{updatee0_t}
\end{aligned} 
\end{equation}

The finite volume method (FVM) is employed to spatially discretize the equation. Namely, by dividing the computational domain into discrete control volumes (cells), and the equation is solved by balancing the energy fluxes across the boundaries of each control volume. The governing equation is integrated over the volume of each cell, transforming the spatial derivative into surface integrals across the cell faces. For a control volume $i$, the discretized form of the BTE becomes,
\begin{equation}
    \sum_{j \in N(i)} \bm{v}_{i,\lambda} \cdot \bm{n}_{ij} S_{ij} e_{ij,\alpha,\lambda} = -\frac{e_{i,\alpha,\lambda} - e_{i,\lambda}^0}{\tau_{i,\lambda}} V_i + Q_{i,\lambda} V_i,
\label{eq:fvm-1}
\end{equation}
where $V_i$ is the volume of the cell $i$ and $N(i)$ is the list of neighboring cells of cell $i$, $S_{ij}$ is the area of the interface between cell $i$ and cell $j$, $\bm{n}_{ij}$ is the outward-facing normal vector at the interface, and $e_{ij,\alpha,\lambda}$ is the phonon energy at the interface, which is computed using upwind scheme to ensure numerical stability,
\begin{equation}
    e_{ij, \alpha,\lambda}=e^\prime +\nabla e^\prime \cdot d^\prime_{ij},
\end{equation}
where $e^\prime$ and  $\nabla e^\prime$ are the phonon energy density and its gradient in the cell from the upwind direction $(\bm{s} \cdot \bm{n} >0)$, while $d^\prime_{ij}$ is the vector from the center of the upwind cell to the center of interface $ij$. The gradient $\nabla e^\prime$ is numerically computed using the Green-Gauss method. 

Directly solving the full system of algebraic equations (Eq.~\ref{eq:fvm-1}) is computationally challenging and unstable due to the presence of the equilibrium energy term, which is a function of the integral of the phonon energy distribution across all directions and bands. To address this, the system is solved iteratively with pseudo-time stepping. Namely, at each iteration (i.e. pseudo-time step) $n+1$, the phonon energy distribution is updated based on the changes (or increments) from the previous iteration $n$. As a result, Eq.~\ref{eq:fvm-1} can be reformulated in its incremental form,
\begin{equation}
    \frac{\Delta e_{i,\alpha,\lambda}^{n+1}}{\tau_{i,\lambda}}+\frac{1}{V_i}\sum_{j\in N\left(i\right)}\Delta e_{ij,\alpha,\lambda}^{n+1}\bm{v}_{i,\lambda}\cdot\bm{n}_{ij}S_{ij}=-\frac{1}{V_i}\sum_{j\in N\left(i\right)}{e_{ij,\alpha,\lambda}^n\bm{v}_{i,\lambda}\cdot\bm{n}_{ij}S_{ij}}-\frac{e_{i,\alpha,\lambda}^n-e_{i,\lambda}^{n,0}}{\tau_{i,\lambda}}+Q_{i,\lambda}
    \label{spatial_eqn}
\end{equation}
where $\Delta e^{n+1}=e^{n+1}-e^n$ is the incremental update of phonon energy density between iterations. This incremental form allows for gradual convergence towards the steady-state solution, ensuring numerical stability and avoiding sharp changes in the solution.

\subsection{Iterative matrix formulation and solution scheme}
Once the element-wise algebraic equations \ref{spatial_eqn} for the discretized BTE have been established, it is advantageous to express these equations in matrix form. Each cell-wise equation can be represented as a linear combination of the unknown phonon energy difference $\Delta e^{n}$, leading to a system of linear equations for all cells under each band-angle combination. This matrix representation facilitates the use of GPU computing, enabling batch processing of the equations and significantly accelerating the simulation. Specifically, the matrix form is expressed as follows:
\begin{equation}
    \bm{A}_{\alpha,\lambda}\Delta e_{\alpha,\lambda\ }^{n+1}=\bm{b}_{\alpha,\lambda}^n,
    \label{Axb}
\end{equation}
where $\bm{A}_{\alpha,\lambda}$ is the stiffness matrix for angle $\alpha$ and band $\lambda$, which has the dimensions of $N_c\times N_c$, with $N_c$ being the total number of cells. The diagonal and off-diagonal elements of the matrix are given by:
\begin{equation}
\begin{aligned}
    A_{ii,\alpha,\lambda} &= \frac{1}{\tau_{i,\lambda}}+\sum_{j\in N\left(i\right)}{\frac{1}{V_i}\bm{v}_{i,\lambda}\cdot}\bm{n}_{ij}H\left(\bm{v}_{i,\lambda}\cdot\bm{n}_{ij}\right)S_{ij}, \\
    A_{ij,\alpha,\lambda} &= \frac{1}{V_i}\bm{v}_{i,\lambda}\cdot\bm{n}_{ij}H\left({-\bm{v}}_{i,\lambda}\cdot\bm{n}_{ij}\right),
    \label{A_matrix}
\end{aligned} 
\end{equation}
where $H(\cdot)$ is the Heaviside step function. The right-side vector $\bm{b}^n$ is expressed as:
\begin{equation}
    \begin{split}
        b_{i,\alpha,\lambda}^n =&-\sum_{j\in N\left(i\right)}{\frac{1}{V_i}\bm{v}_{i,\lambda}\cdot}\bm{n}_{ij}H\left(\bm{v}_{i,\lambda}\cdot\bm{n}_{ij}\right)S_{ij}\left(e_{i,\alpha,\lambda}^n+\nabla e_{i,\alpha,\lambda}^n\cdot\bm{d}_{i,ij}\right) \\
        & -\sum_{j\in N\left(i\right)}{\frac{1}{V_i}\bm{v}_{i,\lambda}\cdot}\bm{n}_{ij}H\left({-\bm{v}}_{i,\lambda}\cdot\bm{n}_{ij}\right)S_{ij}\left(e_{j,\alpha,\lambda}^n+\nabla e_{j,\alpha,\lambda}^n\cdot\bm{d}_{j,ij}\right) \\
        & -\frac{e_{i,\alpha,\lambda}^n}{\tau_{i,\lambda}}+\frac{e_{i,\alpha,\lambda}^{n,0}}{\tau_{i,\lambda}}+Q_{i,\lambda}.
    \end{split}
    \label{rhs_b}
\end{equation}

In these equations, the stiffness matrix $\bm{A}_{\alpha,\lambda}$ depends only on material properties and mesh topology, and thus remains unchanged throughout the simulation process. The vector $\bm{b}^n_{\alpha,\lambda}$, however, needs to be re-evaluated at each iteration after updating the phonon energy density. To solve the linear system, the bi-conjugate gradient (Bi-CG) method is employed, with a Jacobi preconditioner to improve convergence. 

In the iterative solver, the initial phonon energy density is calculated based on the given initial temperature. For each iteration, both the equilibrium energy density and phonon energy density are updated using equation~\ref{update_e0}. The complete solution procedure is summarized as follows~\cite{loy2013fast}:
\begin{enumerate}
    \item 	Set initial equilibrium energy density $e^0$ and phonon energy density $e$ according to the initial temperature.
    \item 	Solve the discretized form of the BTE according to the input boundary conditions to update phonon energy density $e$ according to equation~\ref{spatial_eqn}.
    \item  	Update $e^0$ and $T$ based on equation~\ref{updatee0_t}.
    \item Repeat step 2 and 3 until converge is achieved when the following criteria are satisfied,
    \begin{equation}
    \begin{aligned}
    \epsilon_T &= \sqrt{\sum_{i}^{N}\frac{\left(T_i^n-T_i^{n+1}\right)^2}{N}}/T_{max} < \tilde{\epsilon}_T, \\
    \epsilon_q &= \sqrt{\sum_{i}^{N}\frac{\left(\left|\bm{q}\right|_i^n-\left|\bm{q}\right|_i^{n+1}\right)^2}{N}}/\left|\bm{q}\right|_{max}< \tilde{\epsilon_q},
    \end{aligned} 
    \end{equation}
    where $N$ is the number of spatial cells, $\tilde{\epsilon}_T$ and $\tilde{\epsilon_q}$ are user-defined target temperature residual and heat flux residual, repsectively.
\end{enumerate}

\subsection{Forward and inverse modeling workflow}
\begin{figure}[htp!]
	\centering
    \includegraphics[width=0.9\textwidth]{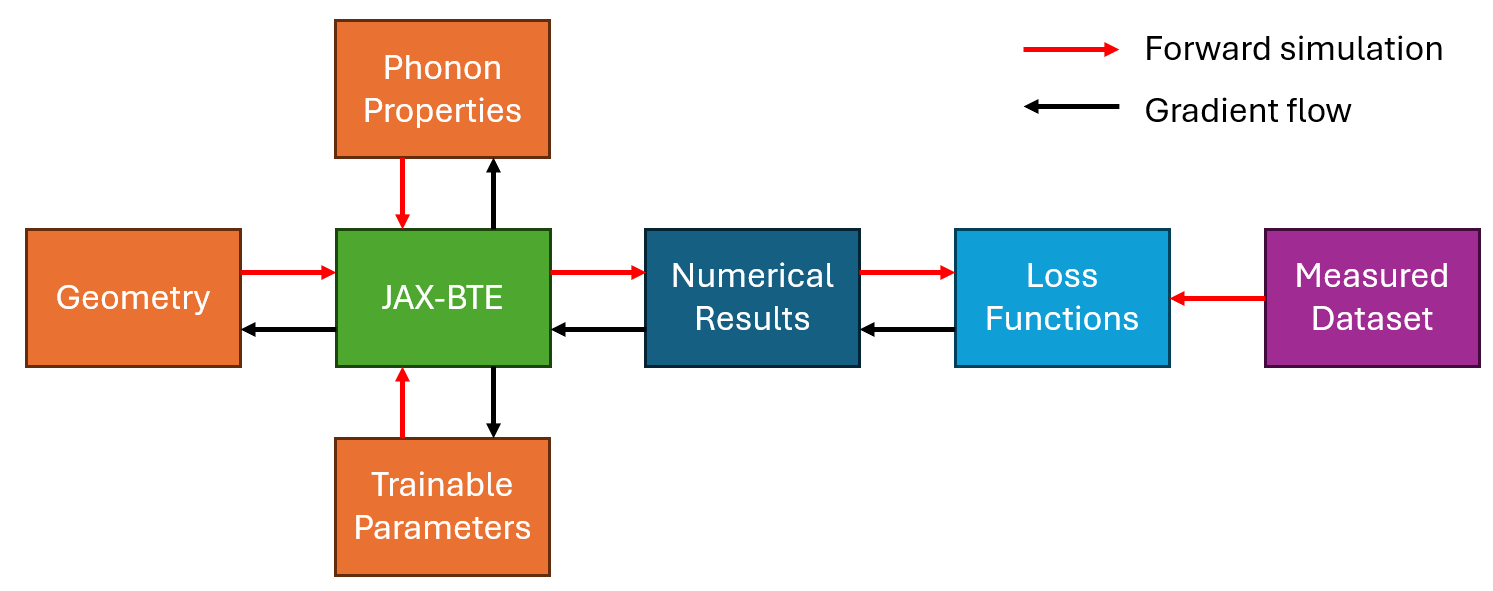}
	\caption{Schematic of the forward and inverse JAX-BTE simulations. Red arrows indicate the forward simulation flow, while black arrows indicate the gradient flow for inverse modeling.}
	\label{fig:workflow}
\end{figure}

The workflow of the JAX-BTE solver for both forward and inverse simulations is illustrated in Figure~\ref{fig:workflow}. The process begins with the geometry configuration, typically represented as a mesh for spatial discretization. The computational mesh, along with material-specific phonon properties, often obtained from first-principles Density Functional Theory (DFT) calculations~\cite{tadano2014anharmonic}, as well as other physical parameters, are fed as the input into the JAX-BTE solver for the forward simulations. Thanks to differentiable programming, inverse simulations can also be conducted if certain input information, such as transistor dimensions or heat source intensity, is unknown while additional measurement data are available. In the inverse setting, the unknowns are defined as trainable parameters, which can be ``trained'' through the gradient-based optimization.

\begin{figure}
	\centering
    \includegraphics[width=0.9\textwidth]{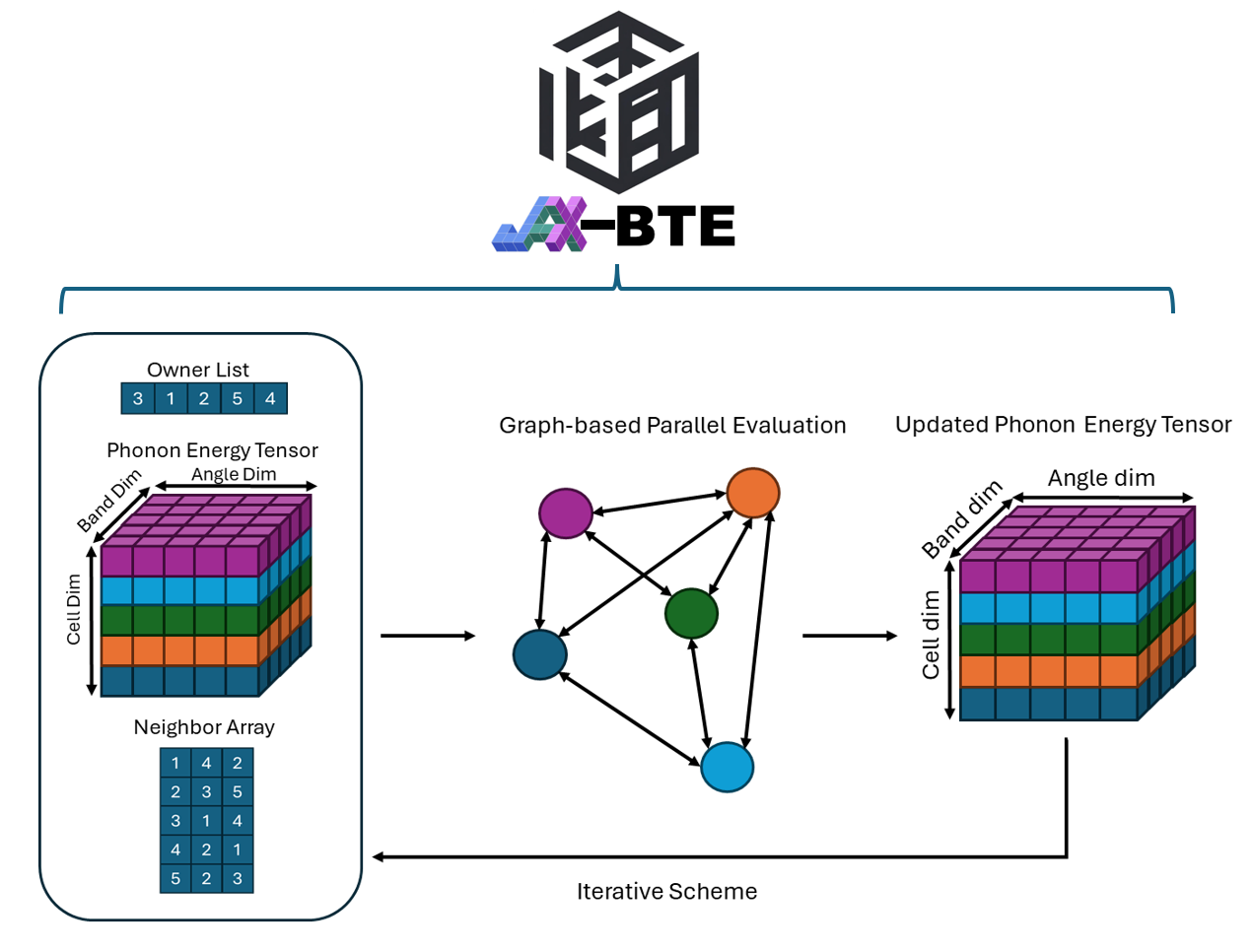}
	\caption{Parallel strategy of the JAX-BTE solver. Phonon energy is stored as a 3D JAX tensor. Owner and neighbor lists are used for the construction of graph-based data structure for efficient GPU computing.} 
	\label{fig:parallel}
\end{figure}

Specifically, the core solver processes the inputs to perform the forward simulation, yielding outputs such as phonon energy distribution, temperature distributions and heat flux. As depicted in Figure~\ref{fig:parallel}, within the solver, phonon energy densities are stored as a three-dimensional JAX tensor, with indices corresponding to the cell index, band index, and angular index. The solver also utilizes the owner and neighbor lists derived from the geometry to map the phonon energy tensors onto a graph that encodes the topological information of the computational mesh. Using this graph-based representation, the solver efficiently computes the phonon energy densities by solving the discretized algebraic equations in parallel using GPUs. This tensor-based parallelization enables efficient updates of the phonon energy distribution across both structured and unstructured meshes.

For inverse modeling tasks, the JAX-BTE solver facilitates the optimization of unknown or uncertain input parameters by comparing simulated results with experimental measurements. These parameters could include material properties, geometric variables, or heat source intensities. The inverse problem is formulated as an optimization problem, where the objective is to minimize a loss function that quantifies the difference between the simulation outputs (e.g., temperature or heat flux) and the corresponding measured data. Given that JAX-BTE is fully differentiable, the loss function can be expressed as a function of the trainable parameters, and its gradient with respect to these parameters is automatically computed through the AD capabilities of JAX. The differentiability feature in JAX-BTE is enabled by the JAX gradient tracking framework. While AD is used for flux calculations and gradient reconstruction, gradient propagation for implicit components, such as the BiCGSTAB solver, is handled through a discrete adjoint approach. Instead of naively applying AD to iterative steps, which would be computationally inefficient, JAX formulates an additional adjoint equation system and solves it using BiCGSTAB. This hybrid strategy leverages the flexibility of AD for explicit computations while utilizing adjoint methods for iterative solvers, ensuring efficient and scalable gradient tracking even for large-scale phonon BTE simulations. This gradient information enables the use of gradient-based optimization algorithms, such as stochastic gradient descent (SGD), Adam, or L-BFGS, to iteratively update the parameters and minimize the loss. The differentiability of the solver ensures that even complex, high-dimensional parameter spaces can be explored systematically, allowing for accurate recovery of unknown parameters or material properties in a physics-constrained setting.

\textbf{Data availability}
All data generated or analysed during this study are included in this published article.

\textbf{Codes availability}
The codes used during the current study are available from the corresponding author on reasonable request.

\textbf{Acknowledgement}
The authors would like to thank DARPA Thermonat Program (HR00112390112) managed by Dr. Yogendra Joshi for the funding support. PD and JXW would also like to acknowledge the funding support from Office of Naval Research under award numbers N00014-23-1-2071. 

\textbf{Author contributions}
TL and JXW contributed to the ideation of the research; WS, YL, JXW designed and programmed the solver. JZ and JP performed mesh generation. WS and JXW wrote the first draft and all authors participated in reviewing and editing the manuscripts.

\textbf{Competing interests}
The authors declare no competing interests.


\begin{thebibliography}{99}
\bibitem{garimella2008thermal}
Garimella, Suresh V and Fleischer, Amy S and Murthy, Jayathi Y and Keshavarzi, Ali and Prasher, Ravi and Patel, Chandrakant and Bhavnani, Sushil H and Venkatasubramanian, Rama and Mahajan, Ravi and Joshi, Yogendra and others, ``Thermal challenges in next-generation electronic systems,'' \textit{IEEE Transactions on Components and Packaging Technologies}, vol. 31, no. 4, pp. 801--815, 2008. IEEE. 
\bibitem{prasher2008cooling}
Prasher, Ravi and Chang, Je-Young, ``Cooling of electronic chips using microchannel and micro-pin fin heat exchangers,'' vol. 48345, pp. 1881--1887, 2008. 
\bibitem{cahill2014nanoscale}
Cahill, David G and Braun, Paul V and Chen, Gang and Clarke, David R and Fan, Shanhui and Goodson, Kenneth E and Keblinski, Pawel and King, William P and Mahan, Gerald D and Majumdar, Arun and others, ``Nanoscale thermal transport. II. 2003--2012,'' \textit{Applied physics reviews}, vol. 1, no. 1, 2014. AIP Publishing.
\bibitem{luo2013nanoscale}
Luo, Tengfei and Chen, Gang, ``Nanoscale heat transfer--from computation to experiment,'' \textit{Physical Chemistry Chemical Physics}, vol. 15, no. 10, pp. 3389--3412, 2013. Royal Society of Chemistry. 
\bibitem{minnich2011quasiballistic}
Minnich, Austin J and Chen, Gang and Mansoor, Saad and Yilbas, BS, ``Quasiballistic heat transfer studied using the frequency-dependent Boltzmann transport equation,'' \textit{Physical Review B—Condensed Matter and Materials Physics}, vol. 84, no. 23, pp. 235207, 2011. APS. 
\bibitem{chen2001phonon}
Chen, Gang, ``Phonon transport in low-dimensional structures,'' vol. 71, pp. 203--259, 2001. Elsevier. 
\bibitem{chen2021non}
Chen, Gang, ``Non-Fourier phonon heat conduction at the microscale and nanoscale,'' \textit{Nature Reviews Physics}, vol. 3, no. 8, pp. 555--569, 2021. Nature Publishing Group UK London. 
\bibitem{cahill2003nanoscale}
Cahill, David G and Ford, Wayne K and Goodson, Kenneth E and Mahan, Gerald D and Majumdar, Arun and Maris, Humphrey J and Merlin, Roberto and Phillpot, Simon R, ``Nanoscale thermal transport,'' \textit{Journal of applied physics}, vol. 93, no. 2, pp. 793--818, 2003. American Institute of Physics. 
\bibitem{toberer2012advances}
Toberer, Eric S and Baranowski, Lauryn L and Dames, Chris, ``Advances in thermal conductivity,'' \textit{Annual Review of Materials Research}, vol. 42, no. 1, pp. 179--209, 2012. Annual Reviews. 
\bibitem{guo2015phonon}
Guo, Yangyu and Wang, Moran, ``Phonon hydrodynamics and its applications in nanoscale heat transport,'' \textit{Physics Reports}, vol. 595, pp. 1--44, 2015. Elsevier. 
\bibitem{minnich2015advances}
Minnich, AJ, ``Advances in the measurement and computation of thermal phonon transport properties,'' \textit{Journal of Physics: Condensed Matter}, vol. 27, no. 5, pp. 053202, 2015. IOP Publishing. 
\bibitem{li2005thermal}
Li, Deyu and Huxtable, Scott T and Abramson, Alexis R and Majumdar, Arun, ``Thermal transport in nanostructured solid-state cooling devices,'' \textit{J. Heat Transfer}, vol. 127, no. 1, pp. 108--114, 2005. 
\bibitem{majumdar2004thermoelectricity}
Majumdar, Arun, ``Thermoelectricity in semiconductor nanostructures,'' \textit{Science}, vol. 303, no. 5659, pp. 777--778, 2004. American Association for the Advancement of Science. 
\bibitem{mazumder2021boltzmann}
Mazumder, Sandip, ``Boltzmann transport equation based modeling of phonon heat conduction: progress and challenges,'' \textit{Annual Review of Heat Transfer}, vol. 24, 2021. Begel House Inc.. 
\bibitem{murthy2005review}
Murthy, Jayathi Y and Narumanchi, Sreekant VJ and Jose'A, Pascual-Gutierrez and Wang, Tianjiao and Ni, Chunjian and Mathur, Sanjay R, ``Review of multiscale simulation in submicron heat transfer,'' \textit{International Journal for Multiscale Computational Engineering}, vol. 3, no. 1, 2005. Begel House Inc.. 
\bibitem{landon2014deviational}
Landon, Colin D and Hadjiconstantinou, Nicolas G, ``Deviational simulation of phonon transport in graphene ribbons with ab initio scattering,'' \textit{Journal of Applied Physics}, vol. 116, no. 16, 2014. AIP Publishing. 
\bibitem{mazumder2001monte}
Mazumder, Sandip and Majumdar, Arunava, ``Monte Carlo study of phonon transport in solid thin films including dispersion and polarization,'' \textit{J. Heat Transfer}, vol. 123, no. 4, pp. 749--759, 2001. 
\bibitem{lacroix2005monte}
Lacroix, David and Joulain, Karl and Lemonnier, Denis, ``Monte Carlo transient phonon transport in silicon and germanium at nanoscales,'' \textit{Physical Review B—Condensed Matter and Materials Physics}, vol. 72, no. 6, pp. 064305, 2005. APS. 
\bibitem{mittal2010monte}
A. Mittal and S. Mazumder, 
``Monte Carlo Study of Phonon Heat Conduction in Silicon Thin Films Including Contributions of Optical Phonons,'' 
\textit{Journal of Heat Transfer}, vol. 132, no. 5, p. 052402, Mar. 2010.
\bibitem{peraud2011efficient}
``Efficient simulation of multidimensional phonon transport using energy-based variance-reduced Monte Carlo formulations,'' \textit{Physical Review B—Condensed Matter and Materials Physics}, vol. 84, no. 20, pp. 205331, 2011. APS.
\bibitem{peraud2015adjoint}
``Adjoint-based deviational Monte Carlo methods for phonon transport calculations,'' \textit{Physical Review B}, vol. 91, no. 23, pp. 235321, 2015. APS. 
\bibitem{escobar2007influence}
R. A. Escobar and C. H. Amon, 
``Influence of Phonon Dispersion on Transient Thermal Response of Silicon-on-Insulator Transistors Under Self-Heating Conditions,'' \textit{Journal of Heat Transfer}, vol. 129, no. 7, pp. 790-797, Sep. 2006. 
\bibitem{escobar2008thin}
R. A. Escobar and C. H. Amon, ``Thin Film Phonon Heat Conduction by the Dispersion Lattice Boltzmann Method,'' 
\textit{Journal of Heat Transfer}, vol. 130, no. 9, p. 092402, Jul. 2008. 
\bibitem{murthy2002computation}
Murthy, JY and Mathur, SR, ``Computation of sub-micron thermal transport using an unstructured finite volume method,'' \textit{J. Heat Transfer}, vol. 124, no. 6, pp. 1176--1181, 2002. 
\bibitem{narumanchi2004submicron}
Narumanchi, Sreekant VJ and Murthy, Jayathi Y and Amon, Cristina H, ``Submicron heat transport model in silicon accounting for phonon dispersion and polarization,'' \textit{J. Heat Transfer}, vol. 126, no. 6, pp. 946--955, 2004. 
\bibitem{ali2014large}
Ali, Syed Ashraf and Kollu, Gautham and Mazumder, Sandip and Sadayappan, P and Mittal, Arpit, ``Large-scale parallel computation of the phonon Boltzmann transport equation,'' \textit{International journal of thermal sciences}, vol. 86, pp. 341--351, 2014. Elsevier. 
\bibitem{sheng2024integrating}
Sheng, Yufei and Wang, Shuying and Hu, Yue and Xu, Jiaxuan and Ji, Zhigang and Bao, Hua, ``Integrating First-principles-based non-Fourier thermal analysis into nanoscale device simulation,'' \textit{IEEE Transactions on Electron Devices}, 2024. IEEE.
\bibitem{saurav2023extraction}
Saurav, Siddharth and Mazumder, Sandip, ``Extraction of thermal conductivity using phonon Boltzmann Transport Equation based simulation of frequency domain thermo-reflectance experiments,'' \textit{International Journal of Heat and Mass Transfer}, vol. 204, pp. 123871, 2023. Elsevier.
\bibitem{saurav2024anisotropic}
Saurav, Siddharth and Mazumder, Sandip, ``Anisotropic Fourier Heat Conduction and phonon Boltzmann transport equation based simulation of time domain thermo-reflectance experiments,'' \textit{International Journal of Heat and Mass Transfer}, vol. 228, pp. 125698, 2024. Elsevier.
\bibitem{bao2018review}
Bao, Hua and Chen, Jie and Gu, Xiaokun and Cao, Bingyang, ``A review of simulation methods in micro/nanoscale heat conduction,'' \textit{ES Energy \& Environment}, vol. 1, no. 84, pp. 16--55, 2018. Engineered Science Publisher. 
\bibitem{zhang2021fast}
Zhang, Chuang and Chen, Songze and Guo, Zhaoli and Wu, Lei, ``A fast synthetic iterative scheme for the stationary phonon Boltzmann transport equation,'' \textit{International Journal of Heat and Mass Transfer}, vol. 174, pp. 121308, 2021. Elsevier. 
\bibitem{hu2022ultra}
Hu, Yue and Shen, Yongxing and Bao, Hua, ``Ultra-efficient and parameter-free computation of submicron thermal transport with phonon Boltzmann transport equation,'' \textit{Fundamental Research}, 2022. Elsevier. 
\bibitem{hu2022optimized}
Hu, Yue and Shen, Yongxing and Bao, Hua, ``Optimized phonon band discretization scheme for efficiently solving the nongray Boltzmann transport equation,'' \textit{Journal of Heat Transfer}, vol. 144, no. 7, pp. 072501, 2022. American Society of Mechanical Engineers. 
\bibitem{karniadakis2021physics}
Karniadakis, George Em and Kevrekidis, Ioannis G and Lu, Lu and Perdikaris, Paris and Wang, Sifan and Yang, Liu, ``Physics-informed machine learning,'' \textit{Nature Reviews Physics}, vol. 3, no. 6, pp. 422--440, 2021. Nature Publishing Group. 
\bibitem{li2021physics}
Li, Ruiyang and Lee, Eungkyu and Luo, Tengfei, ``Physics-informed neural networks for solving multiscale mode-resolved phonon Boltzmann transport equation,'' \textit{Materials Today Physics}, vol. 19, pp. 100429, 2021. Elsevier. 
\bibitem{li2022physics}
Li, Ruiyang and Wang, Jian-Xun and Lee, Eungkyu and Luo, Tengfei, ``Physics-informed deep learning for solving phonon Boltzmann transport equation with large temperature non-equilibrium,'' \textit{npj Computational Materials}, vol. 8, no. 1, pp. 29, 2022. Nature Publishing Group UK London. 
\bibitem{li2023physics}
Li, Ruiyang and Lee, Eungkyu and Luo, Tengfei, ``Physics-informed deep learning for solving coupled electron and phonon Boltzmann transport equations,'' \textit{Physical Review Applied}, vol. 19, no. 6, pp. 064049, 2023. APS. 
\bibitem{zhou2023physics}
Zhou, Jiahang and Li, Ruiyang and Luo, Tengfei, ``Physics-informed neural networks for modeling mesoscale heat transfer using the Boltzmann transport equation,'' vol. 55, pp. 211--238, 2023. Elsevier. 
\bibitem{zhou2023physics_npj}
Zhou, Jiahang and Li, Ruiyang and Luo, Tengfei, ``Physics-informed neural networks for solving time-dependent mode-resolved phonon Boltzmann transport equation,'' \textit{npj Computational Materials}, vol. 9, no. 1, pp. 212, 2023. Nature Publishing Group UK London. 
\bibitem{raissi2019physics}
Raissi, Maziar and Perdikaris, Paris and Karniadakis, George E, ``Physics-informed neural networks: A deep learning framework for solving forward and inverse problems involving nonlinear partial differential equations,'' \textit{Journal of Computational physics}, vol. 378, pp. 686--707, 2019. Elsevier. 
\bibitem{sun2020surrogate}
Sun, Luning and Gao, Han and Pan, Shaowu and Wang, Jian-Xun, ``Surrogate modeling for fluid flows based on physics-constrained deep learning without simulation data,'' \textit{Computer Methods in Applied Mechanics and Engineering}, vol. 361, pp. 112732, 2020. Elsevier. 
\bibitem{cai2021physics}
Cai, Shengze and Wang, Zhicheng and Wang, Sifan and Perdikaris, Paris and Karniadakis, George Em, ``Physics-informed neural networks for heat transfer problems,'' \textit{Journal of Heat Transfer}, vol. 143, no. 6, pp. 060801, 2021. American Society of Mechanical Engineers. 
\bibitem{wang2022and}
Wang, Sifan and Yu, Xinling and Perdikaris, Paris, ``When and why PINNs fail to train: A neural tangent kernel perspective,'' \textit{Journal of Computational Physics}, vol. 449, pp. 110768, 2022. Elsevier. 
\bibitem{gao2021phygeonet}
Gao, Han and Sun, Luning and Wang, Jian-Xun, ``PhyGeoNet: Physics-informed geometry-adaptive convolutional neural networks for solving parameterized steady-state PDEs on irregular domain,'' \textit{Journal of Computational Physics}, vol. 428, pp. 110079, 2021. Elsevier. 
\bibitem{liu2024multi}
Liu, Xin-Yang and Zhu, Min and Lu, Lu and Sun, Hao and Wang, Jian-Xun, ``Multi-resolution partial differential equations preserved learning framework for spatiotemporal dynamics,'' \textit{Communications Physics}, vol. 7, no. 1, pp. 31, 2024. Nature Publishing Group UK London. 
\bibitem{fan2024differentiable}
Fan, Xiantao and Wang, Jian-Xun, ``Differentiable hybrid neural modeling for fluid-structure interaction,'' \textit{Journal of Computational Physics}, vol. 496, pp. 112584, 2024. Elsevier. 
\bibitem{akhare2023diffhybrid}
Akhare, Deepak and Luo, Tengfei and Wang, Jian-Xun, ``Diffhybrid-uq: uncertainty quantification for differentiable hybrid neural modeling,'' \textit{arXiv preprint arXiv:2401.00161}, 2023. 
\bibitem{akhare2023physics}
Akhare, Deepak and Luo, Tengfei and Wang, Jian-Xun, ``Physics-integrated neural differentiable (PiNDiff) model for composites manufacturing,'' \textit{Computer Methods in Applied Mechanics and Engineering}, vol. 406, pp. 115902, 2023. Elsevier. 
\bibitem{fan2024neural}
Fan, Xiantao and Akhare, Deepak and Wang, Jian-Xun, ``Neural Differentiable Modeling with Diffusion-Based Super-resolution for Two-Dimensional Spatiotemporal Turbulence,'' \textit{arXiv preprint arXiv:2406.20047}, 2024. 
\bibitem{akhare2024probabilistic}
Akhare, Deepak and Chen, Zeping and Gulotty, Richard and Luo, Tengfei and Wang, Jian-Xun, ``Probabilistic physics-integrated neural differentiable modeling for isothermal chemical vapor infiltration process,'' \textit{npj Computational Materials}, vol. 10, no. 1, pp. 120, 2024. Nature Publishing Group UK London. 
\bibitem{jax2018github}
J. Bradbury, R. Frostig, P. Hawkins, M. J. Johnson, C. Leary, D. Maclaurin, G. Necula, 
A. Paszke, J. VanderPlas, S. Wanderman-Milne, and Q. Zhang, 
``JAX: composable transformations of Python+NumPy programs,'' 2018. 
\bibitem{frostig2018compiling}
Frostig, Roy and Johnson, Matthew James and Leary, Chris, ``Compiling machine learning programs via high-level tracing,'' \textit{Systems for Machine Learning}, vol. 4, no. 9, 2018. SysML. 
\bibitem{Kochkov2021-ML-CFD}
D. Kochkov, J. A. Smith, A. Alieva, Q. Wang, M. P. Brenner, and S. Hoyer, 
``Machine learning–accelerated computational fluid dynamics,'' 
\textit{Proceedings of the National Academy of Sciences}, 
vol. 118, no. 21, e2101784118, 2021. 
\bibitem{bezgin2024jax}
Bezgin, Deniz A and Buhendwa, Aaron B and Adams, Nikolaus A, ``JAX-Fluids 2.0: Towards HPC for Differentiable CFD of Compressible Two-phase Flows,'' \textit{arXiv preprint arXiv:2402.05193}, 2024. 
\bibitem{schoenholz2020jax}
S. Schoenholz and E. D. Cubuk, 
``JAX MD: a framework for differentiable physics,'' 
\textit{Advances in Neural Information Processing Systems}, 
vol. 33, pp. 11428--11441, 2020.
\bibitem{xue2023jax}
Xue, Tianju and Liao, Shuheng and Gan, Zhengtao and Park, Chanwook and Xie, Xiaoyu and Liu, Wing Kam and Cao, Jian, ``JAX-FEM: A differentiable GPU-accelerated 3D finite element solver for automatic inverse design and mechanistic data science,'' \textit{Computer Physics Communications}, vol. 291, pp. 108802, 2023. Elsevier. 
\bibitem{guo2016discrete}
Guo, Zhaoli and Xu, Kun, ``Discrete unified gas kinetic scheme for multiscale heat transfer based on the phonon Boltzmann transport equation,'' \textit{International Journal of Heat and Mass Transfer}, vol. 102, pp. 944--958, 2016. Elsevier. 
\bibitem{luo2017discrete}
Luo, Xiao-Ping and Yi, Hong-Liang, ``A discrete unified gas kinetic scheme for phonon Boltzmann transport equation accounting for phonon dispersion and polarization,'' \textit{International Journal of Heat and Mass Transfer}, vol. 114, pp. 970--980, 2017. Elsevier. 
\bibitem{sellan2010cross}
Sellan, Daniel P and Turney, JE and McGaughey, Alan JH and Amon, Cristina H, ``Cross-plane phonon transport in thin films,'' \textit{Journal of applied physics}, vol. 108, no. 11, 2010. AIP Publishing. 
\bibitem{maiti2017introducing}
Maiti, Chinmay K, ``Introducing Technology Computer-Aided Design (TCAD): Fundamentals, Simulations, and Applications,'' 2017. Jenny Stanford Publishing. 
\bibitem{hao2018hybrid}
Hao, Qing and Zhao, Hongbo and Xiao, Yue and Wang, Quan and Wang, Xiaoliang, ``Hybrid electrothermal simulation of a 3-D fin-shaped field-effect transistor based on GaN nanowires,'' \textit{IEEE Transactions on Electron Devices}, vol. 65, no. 3, pp. 921--927, 2018. IEEE. 
\bibitem{hu2023giftbte}
Hu, Yue and Jia, Ru and Xu, Jiaxuan and Sheng, Yufei and Wen, Minhua and Lin, James and Shen, Yongxing and Bao, Hua, ``GiftBTE: an efficient deterministic solver for non-gray phonon Boltzmann transport equation,'' \textit{Journal of Physics: Condensed Matter}, vol. 36, no. 2, pp. 025901, 2023. IOP Publishing. 
\bibitem{dames2004theoretical}
Dames, Chris and Chen, Gang, ``Theoretical phonon thermal conductivity of Si/Ge superlattice nanowires,'' \textit{Journal of Applied Physics}, vol. 95, no. 2, pp. 682--693, 2004. American Institute of Physics. 
\bibitem{song2021evaluation}
Song, Qichen and Chen, Gang, ``Evaluation of the diffuse mismatch model for phonon scattering at disordered interfaces,'' \textit{Physical Review B}, vol. 104, no. 8, pp. 085310, 2021. APS. 
\bibitem{ziman2001electrons}
Ziman, John M, ``Electrons and phonons: the theory of transport phenomena in solids,'' 2001. Oxford university press. 
\bibitem{loy2013fast}
Loy, James M and Murthy, Jayathi Y and Singh, Dhruv, ``A fast hybrid Fourier--Boltzmann transport equation solver for nongray phonon transport,'' \textit{Journal of heat transfer}, vol. 135, no. 1, pp. 011008, 2013. American Society of Mechanical Engineers. 
\bibitem{pastrana_jaxfdm_2023}
Bhatnagar, Prabhu Lal and Gross, Eugene P and Krook, Max, ``A model for collision processes in gases. I. Small amplitude processes in charged and neutral one-component systems,'' \textit{Physical review}, vol. 94, no. 3, pp. 511, 1954. APS.
\bibitem{hu2020unification}
Hu, Yue and Feng, Tianli and Gu, Xiaokun and Fan, Zheyong and Wang, Xufeng and Lundstrom, Mark and Shrestha, Som S and Bao, Hua, ``Unification of nonequilibrium molecular dynamics and the mode-resolved phonon Boltzmann equation for thermal transport simulations,'' \textit{Physical Review B}, vol. 101, no. 15, pp. 155308, 2020. APS. 
\bibitem{tadano2014anharmonic}
Tadano, Terumasa and Gohda, Yoshihiro and Tsuneyuki, Shinji, ``Anharmonic force constants extracted from first-principles molecular dynamics: applications to heat transfer simulations,'' \textit{Journal of Physics: Condensed Matter}, vol. 26, no. 22, pp. 225402, 2014. IOP Publishing.

\end{thebibliography}





\end{document}